\documentclass[preprint,review,12pt]{elsarticle}

\usepackage{amssymb} 
\usepackage{amsmath}

\usepackage{graphicx}
\usepackage{array}
\usepackage{booktabs}
\usepackage{graphicx}
\usepackage{subfigure}
\usepackage{multirow}
\usepackage{supertabular}
\usepackage{color}
\usepackage{xcolor}
\usepackage{threeparttable}
\usepackage{algorithm}
\usepackage{algorithmic}
\usepackage{tabularx}

\newcommand{\raa}[1]{\textcolor{black}{#1}}

\journal{Applied Energy}

\begin{document}

\begin{frontmatter}



\title{Spatially Dependent Sampling of Component Failures for Power System Preventive Control Against Hurricane}

\author[tsinghua,keylab]{Ziyue Li} 
\author[tsinghua,keylab]{Guanglun Zhang} 
\author[mit]{Grant Ruan} 
\author[tsinghua,keylab]{Haiwang Zhong\corref{cor1}} 
\author[tsinghua,keylab]{Chongqing Kang} 

\cortext[cor1]{Corresponding author}

\affiliation[tsinghua]{organization={Department of Electrical Engineering, Tsinghua University},
            city={Beijing},
            postcode={100084}, 
            country={China}}
\affiliation[keylab]{organization={State Key Laboratory of Power System Operation and Control},
            city={Beijing},
            postcode={100084}, 
            country={China}}
\affiliation[mit]{organization={Laboratory for Information and Decision Systems (LIDS), Massachusetts Institute of Technology},
            city={Boston},
            postcode={02139}, 
            state={MA},
            country={United States}}

\begin{abstract}
Preventive control is a crucial strategy for power system operation against impending natural hazards, and its effectiveness fundamentally relies on the realism of scenario generation. While most existing studies employ sequential Monte Carlo simulation and assume independent sampling of component failures, this oversimplification neglects the spatial correlations induced by meteorological factors such as hurricanes. In this paper, we identify and address the gap in modeling spatial dependence among component failures under extreme weather. We analyze how the mean, variance, and correlation structure of weather intensity random variables influence the correlation of component failures. To fill this gap, we propose a spatially dependent sampling method that enables joint sampling of multiple component failures by generating correlated meteorological intensity random variables. Comparative studies show that our approach captures long-tailed scenarios and reveals more extreme events than conventional methods. Furthermore, we evaluate the impact of scenario selection on preventive control performance. Our key findings are: (1) Strong spatial correlations in uncertain weather intensity consistently lead to interdependent component failures, regardless of mean value level; (2) The proposed method uncovers more high-severity scenarios that are missed by independent sampling; (3) Preventive control requires balancing load curtailment and over-generation costs under different scenario severities; (4) Ignoring failure correlations results in underestimating risk from high-severity events, undermining the robustness of preventive control strategies.
\end{abstract}


\begin{highlights}
\item Correlation in weather intensity induces correlation in component failures.
\item A spatially dependent sampling approach is developed to capture failure interdependence.
\item Preventive control balances load curtailment and over-generation costs.
\item Neglecting failure interdependence prevents robust preventive control.
\end{highlights}

\begin{keyword}
preventive control \sep
unit commitment \sep 
extreme weather \sep
spatially dependent sampling



\end{keyword}

\end{frontmatter}


\section{Introduction}

\subsection{Background}

As power systems expand and smart grid technologies advance, ensuring their secure and stable operation has become an increasingly critical challenge. However, this development also exposes such systems to external factors, particularly weather conditions, that significantly affect operational performance and component reliability~\cite{WeatherFactor}.

Environmental concerns, especially extreme weather events, have received growing attention across industries. The increasing frequency of such events poses substantial threats to power system security and results in significant societal losses. For instance, Hurricane Sandy in 2013 caused 147 fatalities and left 8.5 million people without power for extended periods~\cite{SandyReport}. Similarly, Typhoon Mangkhut in 2018 led to widespread power outages and economic losses exceeding 22 billion USD~\cite{Mangkhut}, while Hurricane \textit{Harvey} in 2017 left 336,000 people without electricity~\cite{Harvey}.

Adverse weather events, such as storms and hail, tend to follow specific paths, affecting power system components and potentially triggering cascading failures. Therefore, it is essential to develop control strategies tailored to specific meteorological conditions. High-accuracy forecasts produced by Numerical Weather Prediction (NWP) systems enable proactive planning for such events. By leveraging these forecasts, preventive control strategies can effectively mitigate the adverse impacts of weather on power systems.

\subsection{Preventive Control Methods}

\begin{figure}
	\centering
	{\includegraphics[width=0.9\linewidth]{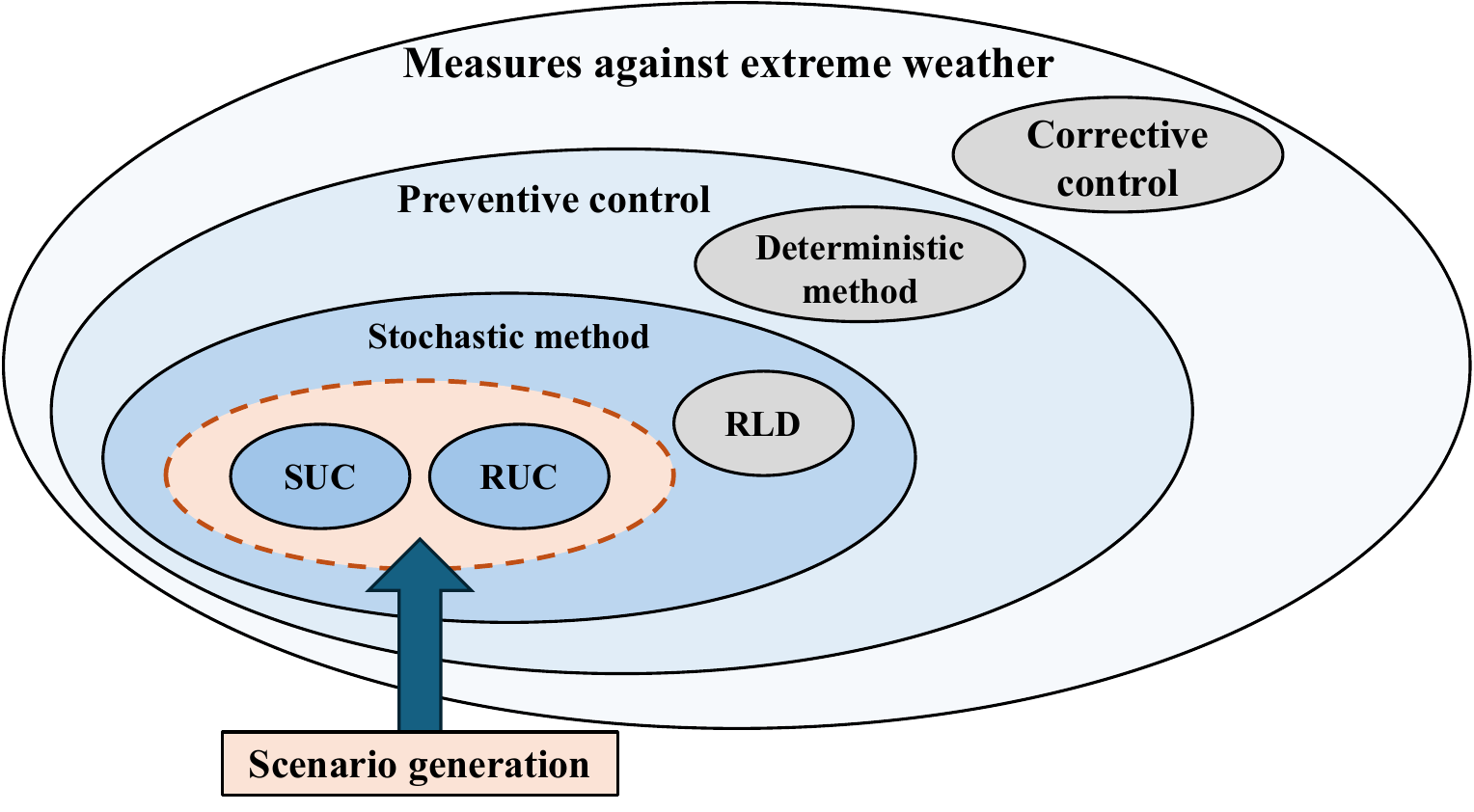}}
	\caption{Structure of literature review on preventive control methods}
	\label{fig:structure_of_literature_review}
\end{figure}

To enhance robustness against extreme weather, power systems must improve their resilience~\cite{BattleExtremeResilience, ResilientHeYutong}. Research in this area can be broadly categorized into two approaches: corrective control strategies and preventive control strategies. Corrective control focuses on response measures after contingencies occur, such as fault diagnosis, emergency response planning, and load recovery~\cite{FaultDiagnosis,DisposalPlan,LoadRecovery}. In contrast, preventive control aims to address potential faults or impending contingencies before they materialize by adjusting operational parameters, such as unit commitment.

Current research on preventive control primarily adopts two methodological frameworks: deterministic and stochastic approaches. Deterministic methods, such as the N-1 and N-k security criteria, treat all possible risks equally, rather than focusing solely on the most probable events. These approaches are typically conservative and inefficient~\cite{N-kCriterion, RobustN-k}. In contrast, stochastic methods emphasize the most likely scenarios by explicitly modeling uncertainties, as seen in risk-limiting dispatch (RLD), stochastic unit commitment (SUC), and robust unit commitment (RUC).

RLD models uncertainty as risk and introduces chance constraints. It was first introduced in~\cite{RLD} and later extended in~\cite{RLDOPCons} to include additional operational constraints. In~\cite{RiskLimitingUC}, a risk-limiting unit commitment model is proposed, where the risk of loss of load (LOL) is quantified using conditional value at risk (CVaR).

SUC and RUC address uncertainty through scenario-based modeling. SUC incorporates a set of possible scenarios into the unit commitment model, and is typically solved by directly handling the resulting scenario-based optimization problem or using decomposition and coordination algorithms. Various studies~\cite{MLSUC, SUC3stage, TSUC, MEGSUC, SequenceProactiveControl, StochasticPreHurricance} have proposed SUC-based preventive control models for hurricanes. For example,~\cite{TSUC} applies unit commitment for preventive control, while~\cite{MLSUC} accelerates the solution process using data-driven techniques. In~\cite{MEGSUC}, mobile emergency generators are pre-deployed to mitigate hurricane impacts. A long-term SUC model to estimate the cost of maintaining system reliability is proposed in~\cite{CostStochasticUnitCommitment}, where generator and transmission line outages are modeled via scenario trees using Monte Carlo simulations. A decomposition algorithm for multi-stage SUC problems is presented in~\cite{SUCDualDecomposition}, in which unit commitment decisions are made in the first stage, prior to uncertainty realization in the second stage.

RUC, which focuses on worst-case scenarios, can be considered as a special case of SUC. In~\cite{MEGRUC}, mobile energy resources are scheduled before extreme weather events using a robust optimization model. Reference~\cite{RobustN-k} proposes a robust N-k security-constrained optimal power flow model capable of handling up to three transmission line contingencies in a medium-sized system. In~\cite{RCUC}, the worst-case scenario is calculated using a duality-based subproblem, although this approach may encounter computational limitations in large-scale systems, where only near-worst-case scenarios may be feasible. 
\raa{For large scale systems, it might be infeasible to found the exact-worst-case scenario before extreme weather event arrives due to computation time. In real-word preventive control, it would be more practical to use heuristic methods (e.g. index-based methods) to obtain near-worst-case scenarios quickly instead. Thus more valuable time before extreme weather event can be left for the preventive control decision making.}

Preventive control is a more efficient strategy for managing foreseeable extreme weather events. 
\raa{Since it utilizes more information about future operating states to make decisions, enabling more targeted allocation of resources. Decisions made by deterministic methods may require higher costs for reallocating resources or incur penalty losses such as load shedding in future fault scenarios.}
Among these preventive control methods, stochastic approaches are particularly effective in achieving optimal scheduling. SUC- and RUC-based methods are the most commonly adopted, and scenario generation plays a crucial role in their performance. The accuracy of scenario generation directly determines the effectiveness of preventive control. However, there remain open problems in generating realistic failure scenarios under extreme weather.

\subsection{Scenario Generation Challenges and Our Contribution: Spatially Dependent Sampling}

\begin{figure}
	\centering
	\includegraphics[width=1.0\linewidth]{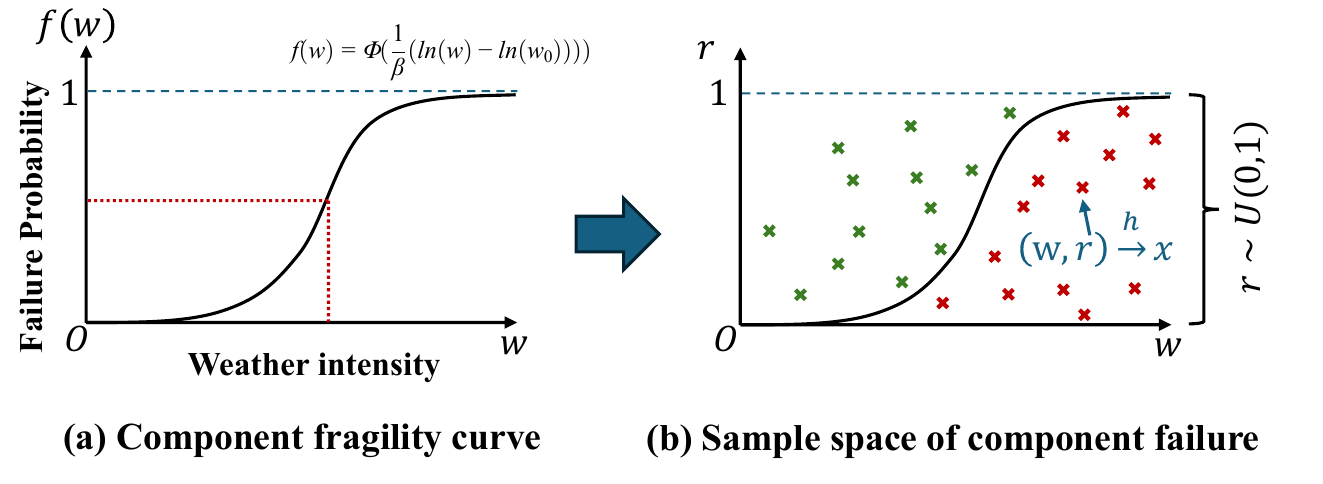}
	\caption{Fragility curve and sample space for single component failure}
	\label{fig:single_component_failure_sampling}
\end{figure}

To simulate the failure of a single component, it is commonly assumed that its resistance to natural hazards follows a lognormal distribution~\cite{FragilityModelling}. The failure probability under a given weather intensity can be described by the fragility function:
\begin{align}
	P(x=1|w) = f(w) = \Phi \left(\frac{1}{\beta}(\ln(w)-\ln(w_0))\right)
\end{align}
where $x$ indicates whether failure occurs and $w$ is the weather intensity. Parameters $\beta$ and $w_0$ represent the component's fragility characteristics, and $\Phi(\cdot)$ is the cumulative distribution function of the standard normal distribution. This can also be expressed as a function of the sample space. The random variable $x$ is represented as a function $h(w,r)$, where $r$ is a uniform random variable $U(0,1)$. Figure~\ref{fig:single_component_failure_sampling} illustrates the fragility curve and corresponding sample space.

\begin{figure}
	\centering
	\includegraphics[width=0.9\linewidth]{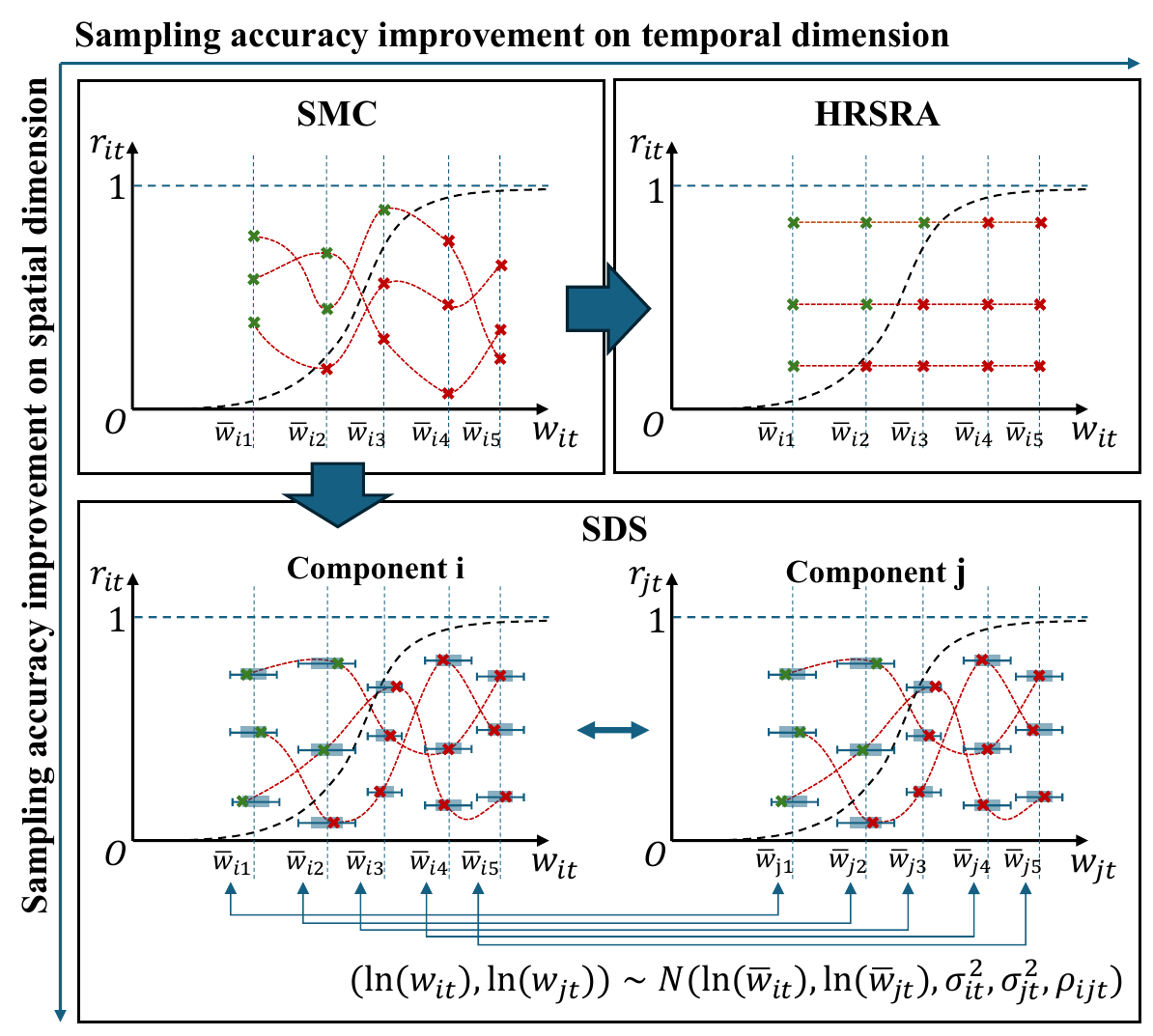}
	\caption{\raa{Comparison of different sampling methods in the sample space}}
	\label{fig:sampling_method_comparison}
\end{figure}

A power system failure scenario generally comprises multiple component failures across various time intervals, thereby extending the single-component failure model across temporal and spatial dimensions. Figure~\ref{fig:sampling_method_comparison} compares different sampling methods in the sample space (for illustrative purposes, assume weather intensity increases over time). Each dotted line represents the failure sequence of a component within one scenario.

Most previous works extend the single-component model to multi-component, multi-period scenarios simply by sampling each component's failure independently in each period, often using sequential Monte Carlo (SMC) methods. However, this approach systematically overlooks the impact of temporal and spatial correlation among failures, which can lead to significant misestimation of system risk.

On the temporal dimension,~\cite{HRSRA} recently showed that SMC methods tend to overestimate failure risk as temporal resolution increases, and proposed a method to mitigate this \raa{temporal inconsistency} by sharing random variable $r$ across time steps. 
\raa{In this way, scenarios generated with high temporal resolution and low temporal resolution are consistent.}

On the spatial dimension, different components may be affected by the same uncertain meteorological conditions. However, the correlation between their failures has not been well studied. Ignoring spatial dependence may cause underestimation of the risk of large-scale, simultaneous failures, especially in extreme events.

To fill this gap, we propose a spatially dependent sampling (SDS) method that explicitly considers the joint distribution of weather intensities for all components exposed to an extreme event (e.g., a hurricane). Our approach generates realistic, spatially correlated failure scenarios, overcoming the limitations of conventional independent-sampling SMC methods.

The main contributions of this paper are as follows:
\begin{enumerate}
	\item We identify and address the spatial correlation gap in failure scenario generation under extreme weather, proposing an SDS technique that accurately simulates correlated multi-component failures.
	\item We systematically compare SDS and conventional SMC scenario sets using a synthetic Texas grid and simulated Hurricane \textit{Harvey} data, revealing the risk underestimation inherent to independent sampling.
	\item We formulate a scenario-based preventive unit commitment model, demonstrating that considering spatial correlation significantly improves the robustness of preventive scheduling for extreme weather events.
\end{enumerate}

The remainder of this paper is organized as follows: Section~\ref{sec:two_component_failure} discusses the correlation of component failures under correlated weather intensities. Section~\ref{sec:relevance_sampling_technique} introduces the spatially dependent sampling method. Section~\ref{sec:preventive_control_method} outlines the preventive control process and formulates a stochastic unit commitment model. Section~\ref{sec:case_study} presents a case study using the synthetic Texas grid and simulated Hurricane \textit{Harvey} to evaluate the performance of the proposed SDS method relative to standard SMC approaches.

\section{Correlation Between Two Component Failures Under Correlated Weather Intensity}
\label{sec:two_component_failure}

In this section, we analyze the scenario involving only two components, which can be any pair within the affected range of a hurricane. Our objective is to determine the extent to which correlated weather intensities can induce correlation between component failures.

\subsection{Failure Sampling Model Based on Relative Weather Intensity}

\raa{
We assume that the weather intensities $w_i, w_j$ at components $i$ and $j$ follow a joint log-normal distribution. 
This is because uncertain weather intensity $w_i$ is the result of the combined effect of multiple random variables (i.e. parameters in the weather field model), and such nonlinear effects are approximately ``multiplicative". According to the Central Limit Theorem, the distribution of the uncertain weather intensity $w_i$ should be approximately log-normal.
And random variables $r_i$ and $r_j$ follow uniform distributions. 
}
\begin{align}
	(\ln{w_i},\ln{w_j}) \sim N\left(\ln{\overline{w}_i}, \ln{\overline{w}_j}, \sigma_i, \sigma_j, \rho_{ij}\right)
	\label{eq:two_component_lnw_dist}
\end{align}
\begin{align}
	r_i, r_j \sim U(0,1)
	\label{eq:two_component_r_dist}
\end{align}

Each component has a fragility function $f_i(w_i)$, and the failure state $x_i$ is defined as a function $h(w_i, r_i)$ of $w_i$ and $r_i$:
\begin{align}
	f_i(w_i) = \Phi\left(\frac{1}{\beta_i} (\ln{w_i} - \ln{w_{i0}})\right)
	\label{eq:two_component_fragility_function}
\end{align}
\begin{align}
	x_i = h(w_i, r_i) = \begin{cases}
		1, & \text{if } r_i < f(w_i) \\
		0, & \text{otherwise}
	\end{cases}
	\label{eq:two_component_h_function}
\end{align}

There are nine parameters in total: two fragility parameters ($\beta_i$, $w_{i0}$) per component, two weather parameters ($\ln \overline{w}_i$, $\sigma_i$) per component, and one correlation coefficient $\rho_{ij}$ between weather intensities. To reduce the degrees of freedom and derive more general conclusions irrespective of specific component parameters, we normalize the weather intensity variables relative to the fragility parameters:
\begin{align}
	w^*_i = \frac{\ln{w_i} - \ln{w_{i0}}}{\beta_i}
\end{align}
\begin{align}
	\overline{w}^*_i = \frac{\ln{\overline{w}_i} - \ln{w_{i0}}}{\beta_i}
\end{align}
\begin{align}
	\sigma^*_i = \frac{\sigma_i}{\beta_i}
\end{align}

Accordingly, the sampling model becomes:
\begin{align}
	(w^*_i, w^*_j) \sim N\left(\overline{w}^*_i, \overline{w}^*_j, \sigma^*_i, \sigma^*_j, \rho_{ij} \right)
\end{align}
\begin{align}
	f^*(w^*_i) = \Phi(w^*_i)
\end{align}
\begin{align}
	x_i = h^*_i(w^*_i, r_i) = \begin{cases}
		1, & \text{if } r_i < f^*(w^*_i) \\
		0, & \text{otherwise}
	\end{cases}
\end{align}

\begin{figure}
	\centering
	{\includegraphics[width=0.6\linewidth]{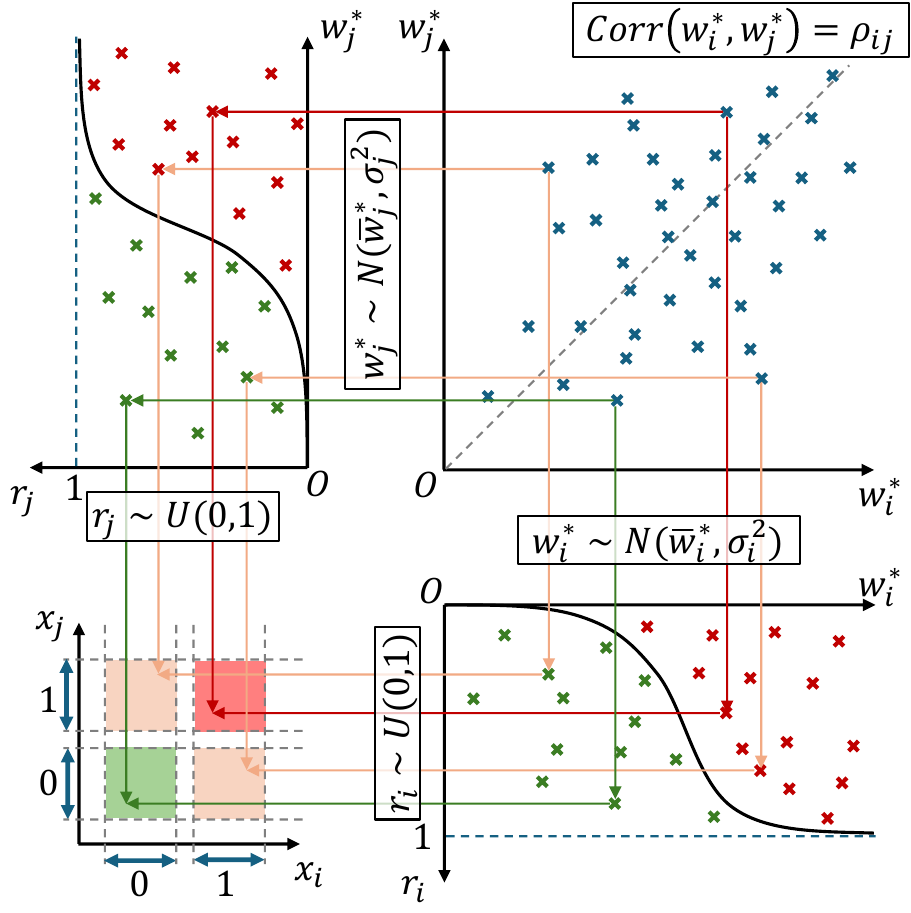}}
	\caption{Sampling process of two components under correlated weather intensity}
	\label{fig:two_component_sampling_structure}
\end{figure}

The sampling process is illustrated in Figure~\ref{fig:two_component_sampling_structure}. The key step is the simultaneous generation of correlated normalized weather intensities $w^*_i$ and $w^*_j$.

\subsection{Sensitivity Analysis of Weather Intensity on Component Failure Correlation}
\begin{figure}
	\centering
	{\includegraphics[width=0.95\linewidth]{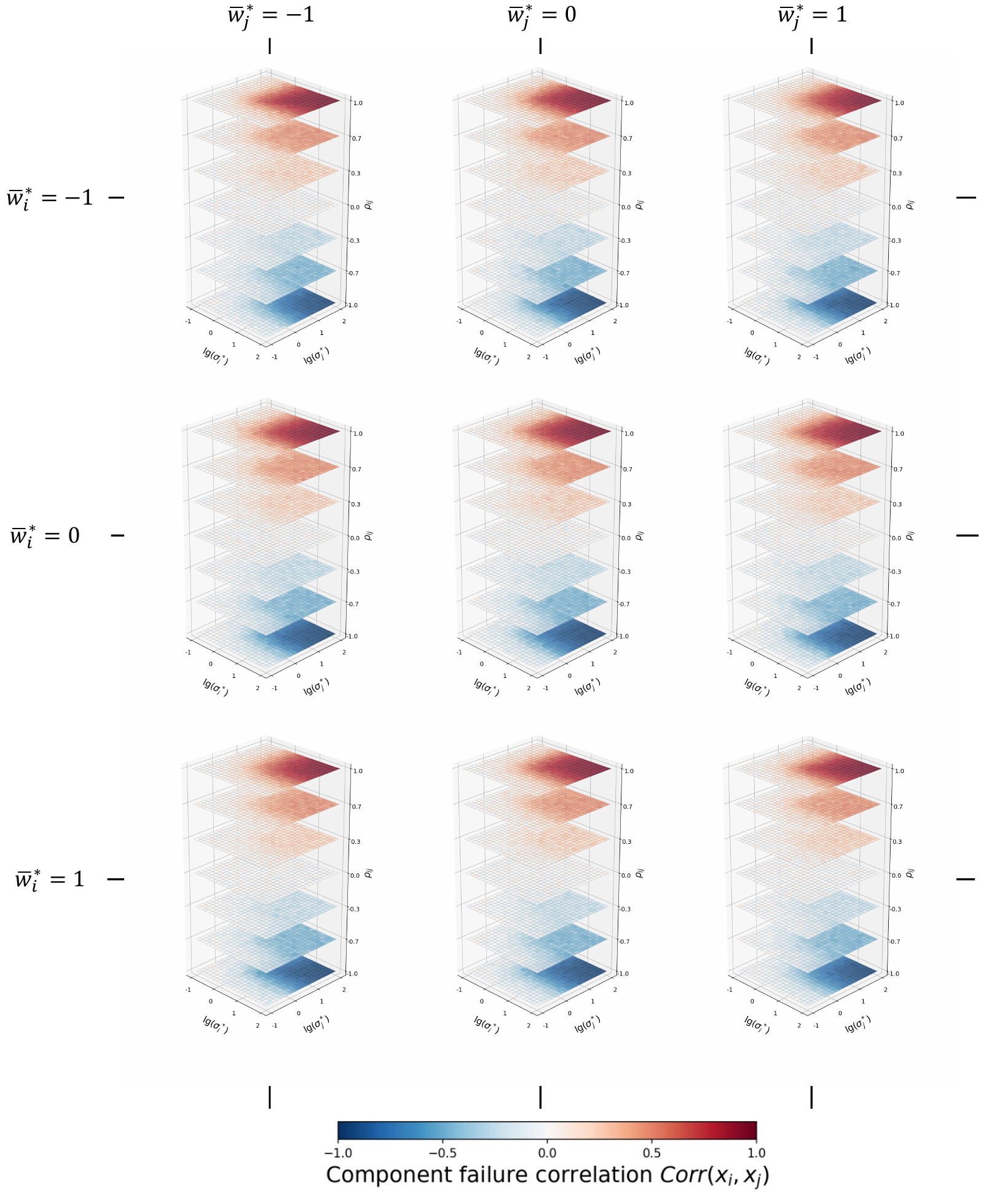}}
	\caption{\raa{Sensitivity of $\mathrm{Corr}(x_i,x_j)$ to $\overline{w}^*_i$, $\overline{w}^*_j$, $\sigma^*_i$, $\sigma^*_j$, and $\rho_{ij}$.}}
	\label{fig:two_component_sensitivity}
\end{figure}

\raa{
We study how the mean values $\overline{w}^*_i$, $\overline{w}^*_j$, the standard deviations $\sigma^*_i$, $\sigma^*_j$, and the correlation coefficient $\rho_{ij}$ affect the correlation between $x_i$ and $x_j$.
First, we fix $\overline{w}^*_i \in \{-1,0,1\}$ and $\overline{w}^*_j \in \{-1,0,1\}$.
We then vary $\sigma^*_i$ and $\sigma^*_j$ on a logarithmic grid from $10^{-1}$ to $10^{2}$, and set $\rho_{ij} \in \{-1,-\tfrac{2}{3},-\tfrac{1}{3},0,\tfrac{1}{3},\tfrac{2}{3},1\}$.
For each parameter set, we generate 3{,}000 scenarios and compute $\mathrm{Corr}(x_i,x_j)$.
The results are shown in Figure~\ref{fig:two_component_sensitivity}.
}

\raa{
The results show that $\rho_{ij}$ is the primary driver of $\mathrm{Corr}(x_i,x_j)$.
When $\rho_{ij}=0$, we observe no correlation between $x_i$ and $x_j$.
Larger $\sigma^*_i$ and $\sigma^*_j$ amplify the induced correlation.
Moreover, the effect of $\sigma^*_i$ and $\sigma^*_j$ strengthens as $|\rho_{ij}|$ increases; for a given pair $(\sigma^*_i,\sigma^*_j)$, a larger $|\rho_{ij}|$ yields higher $\mathrm{Corr}(x_i,x_j)$.
}

\raa{
The mean weather intensities $\overline{w}^*_i$ and $\overline{w}^*_j$ have only a minor effect.
Across the nine tested combinations of $(\overline{w}^*_i,\overline{w}^*_j)$, the patterns are nearly identical as $\sigma^*_i$, $\sigma^*_j$, and $\rho_{ij}$ vary.
This suggests that uncertain and correlated weather intensities can induce component failure correlation regardless of the mean intensity level.
Given the limits of forecasting, it is reasonable to treat upcoming weather intensities as random variables.
We discuss their cross-location correlation in the next section.
}

\section{Spatially Dependent Sampling}
\label{sec:relevance_sampling_technique}

In this section, we introduce the spatially dependent sampling technique for multiple components. We first propose a straightforward method to estimate the correlation between weather intensities on each component. Based on this method, we then describe the comprehensive process of spatially dependent sampling.

\subsection{Estimation of Weather Intensity Correlation}

To extend the two-component scenario to a multi-component framework, the parameter $\rho_{ij}$ must be expanded into a covariance matrix $\mathbf{C}$. Below, we propose a simple method for estimating a postive semi-definite covariance matrix.

Assume we have a hurricane wind field model with parameters $\theta_{kt}$, where $k \in \mathbf{K}$, at time $t$. The weather intensity at each component $i$ can be expressed as a function:
\begin{align}
	w_{it} = f^W_i(\theta_{1,t}, \theta_{2,t}, \dots, \theta_{|\mathbf{K}|,t})
\end{align}

Due to the imperfect prediction of each hurricane parameter, each $\theta_{kt}$ is a random variable with variance:
\begin{align}
	\text{Var}(\theta_{kt}) = \sigma^2_{kt}
\end{align}

We linearize $\ln{w_{it}}$ around its predicted value:
\begin{align}
	\ln{w_{it}} \approx \ln{\overline{w}_{it}} + \sum_{k \in \mathbf{K}} \frac{\partial f^W_i(\overline{\theta}_{1,t}, \overline{\theta}_{2,t}, \dots, \overline{\theta}_{nt})}{\overline{w}_{it} \partial \theta_{kt}} (\theta_{kt} - \overline{\theta}_{kt})
\end{align}

Using this linearization, the covariance between each pair of $\ln{w_{it}}$ and $\ln{w_{jt}}$ can be computed as:
\begin{align}
	&\mathrm{Cov}(\ln{w_{it}},\ln{w_{jt}})= \\\notag
	&\sum_{k\in\mathbf{K}}
	\frac{\partial f^W_i(\overline{\theta}_{1,t},\overline{\theta}_{2,t},...,\overline{\theta}_{nt})}{\overline{w}_{it}\partial \theta_{kt}}
	\frac{\partial f^W_j(\overline{\theta}_{1,t},\overline{\theta}_{2,t},...,\overline{\theta}_{nt})}{\overline{w}_{jt}\partial \theta_{kt}}
	\sigma^2_{kt}
	\label{}
\end{align}
This equation shows that the correlation between weather intensities depends on how they are influenced by the same factors and the uncertainty associated with those factors.

\raa{
Our modeling chain is: parameter uncertainty $\sigma^2_{kt}\ \Rightarrow$ uncertainty and spatial correlation of weather intensity $w_{it}\ \Rightarrow$ correlation of component failures $x_{it}$. (i) If we \emph{uniformly scale} all parameter variances by a factor $c>1$, the marginal spread of weather intensity at every location increases, while the \emph{pairwise correlations of} $\ln w_{it}$ remain essentially unchanged (numerator and denominator scale together). According to Section~\ref{sec:two_component_failure}, this larger spread tends to push the failure correlation $\mathrm{Corr}(x_{it},x_{jt})$ \emph{further away from zero} (it increases when positive and decreases when negative). (ii) If we \emph{increase only one parameter's variance} $\sigma^2_{k^\star t}$, the variability at all locations grows, but the covariance between sites $i$ and $j$ shifts according to the \emph{signs} of their sensitivities to that parameter: when the sensitivities share the same sign, the covariance increases; when the signs differ, the covariance decreases. Because baseline covariances can be of either sign, the net change in $\mathrm{Corr}(w_{it},w_{jt})$, and thus in $\mathrm{Corr}(x_{it},x_{jt})$ is model- and location-dependent. These qualitative effects clarify how stronger (weaker) forecast uncertainty generally widens (narrows) weather intensity spreads and correspondingly strengthens (weakens) the induced failure dependence, while targeted increases in a single parameter's uncertainty can either reinforce or offset spatial co-variation depending on sensitivity alignment.
}

For all the components $\mathbf{\tilde{L}}$ considered, we compute the partial derivative vector $\mathbf{V}_{kt}(\mathbf{\tilde{L}})$ with respect to each parameter $\theta_{kt}$:
\begin{align}
	\mathbf{V}_{kt}(\mathbf{\tilde{L}}) =
	\left[
	\begin{array}{c}
	\frac{1}{\overline{w}_{1,t}} \cdot \frac{\partial f^W_1(\overline{\theta}_{1,t}, \overline{\theta}_{2,t}, \dots, \overline{\theta}_{nt})}{\partial \theta_{kt}} \\
	\frac{1}{\overline{w}_{2,t}} \cdot \frac{\partial f^W_2(\overline{\theta}_{1,t}, \overline{\theta}_{2,t}, \dots, \overline{\theta}_{nt})}{\partial \theta_{kt}} \\
	\vdots \\
	\frac{1}{\overline{w}_{|\mathbf{\tilde{L}}|,t}} \cdot \frac{\partial f^W_{|\mathbf{\tilde{L}}|}(\overline{\theta}_{1,t}, \overline{\theta}_{2,t}, \dots, \overline{\theta}_{nt})}{\partial \theta_{kt}}
	\end{array}
	\right]
\end{align}

Then, the covariance matrix $\mathbf{C}_t(\mathbf{\tilde{L}})$ for the weather intensities on components $\mathbf{\tilde{L}}$ at time $t$ can be calculated as:
\begin{align}
	\mathbf{C}_t(\mathbf{\tilde{L}}) = \sum_{k \in \mathbf{K}} \sigma^2_{kt} \mathbf{V}_{kt}(\mathbf{\tilde{L}}) \mathbf{V}_{kt}^T(\mathbf{\tilde{L}})
	\label{eq:Ct_calculation}
\end{align}
This matrix $\mathbf{C}_t$ is guaranteed to be postive semi-definite, as required by the joint normal distribution.

\raa{
In this study we set cross-parameter covariances to zero, that is, $\mathrm{Cov}(\theta_{k,t},\theta_{\ell,t})=0$ for $k\neq \ell$, so that the propagated covariance in \eqref{eq:Ct_calculation} is a transparent sum of rank-1 contributions $\sigma_{kt}^2\,\mathbf{V}_{kt}(\mathbf{\tilde{L}})\mathbf{V}_{kt}^T(\mathbf{\tilde{L}})$ and is positive semi-definite by construction. We acknowledge that correlations among parameters may exist, but reliable joint error statistics are typically unavailable compared with marginal skill measures (e.g., RMSEs), and attempting to infer off-diagonal terms from limited data can lead to unstable or weakly identified calibrations. Our independence assumption therefore reduces calibration burden while preserving the core mechanism captured above: two locations are more strongly correlated when they significantly respond to the same parameter and that parameter is highly uncertain. When credible cross-parameter information becomes available, the formula naturally generalizes to
\begin{align}
	\mathbf{C}_t(\mathbf{\tilde{L}})=\sum_{k,\ell\in\mathbf{K}}\mathrm{Cov}(\theta_{k,t},\theta_{\ell,t})\,\mathbf{V}_{kt}(\mathbf{\tilde{L}})\mathbf{V}_{\ell t}^T(\mathbf{\tilde{L}})
	=V_t(\mathbf{\tilde{L}})\,\mathbf{\Sigma}_{\theta,t}\,V_t^T(\mathbf{\tilde{L}})
	\label{eq:Ct_aug_calculation}
\end{align}
where $V_t(\mathbf{\tilde{L}})=[\,\mathbf{V}_{1t}(\mathbf{\tilde{L}})\ \cdots\ \mathbf{V}_{Kt}(\mathbf{\tilde{L}})\,]$ and $\mathbf{\Sigma}_{\theta,t}$ collects the parameter covariances. 
This expression is positive semi-definite whenever $\mathbf{\Sigma}_{\theta,t}$ is. Practitioners who have cross-parameter covariance information, or who use advanced physical or AI weather models that directly synthesize correlated weather intensities (especially for hurricanes), can use our pipeline. They may provide either the resulting $\mathbf{\Sigma}_{\theta,t}$ or joint weather intensity samples. The pipeline will then perform joint sampling of component failures.
}

\subsection{Spatially Dependent Sampling Process}

\begin{figure}
	\centering
	{\includegraphics[width=1.0\linewidth]{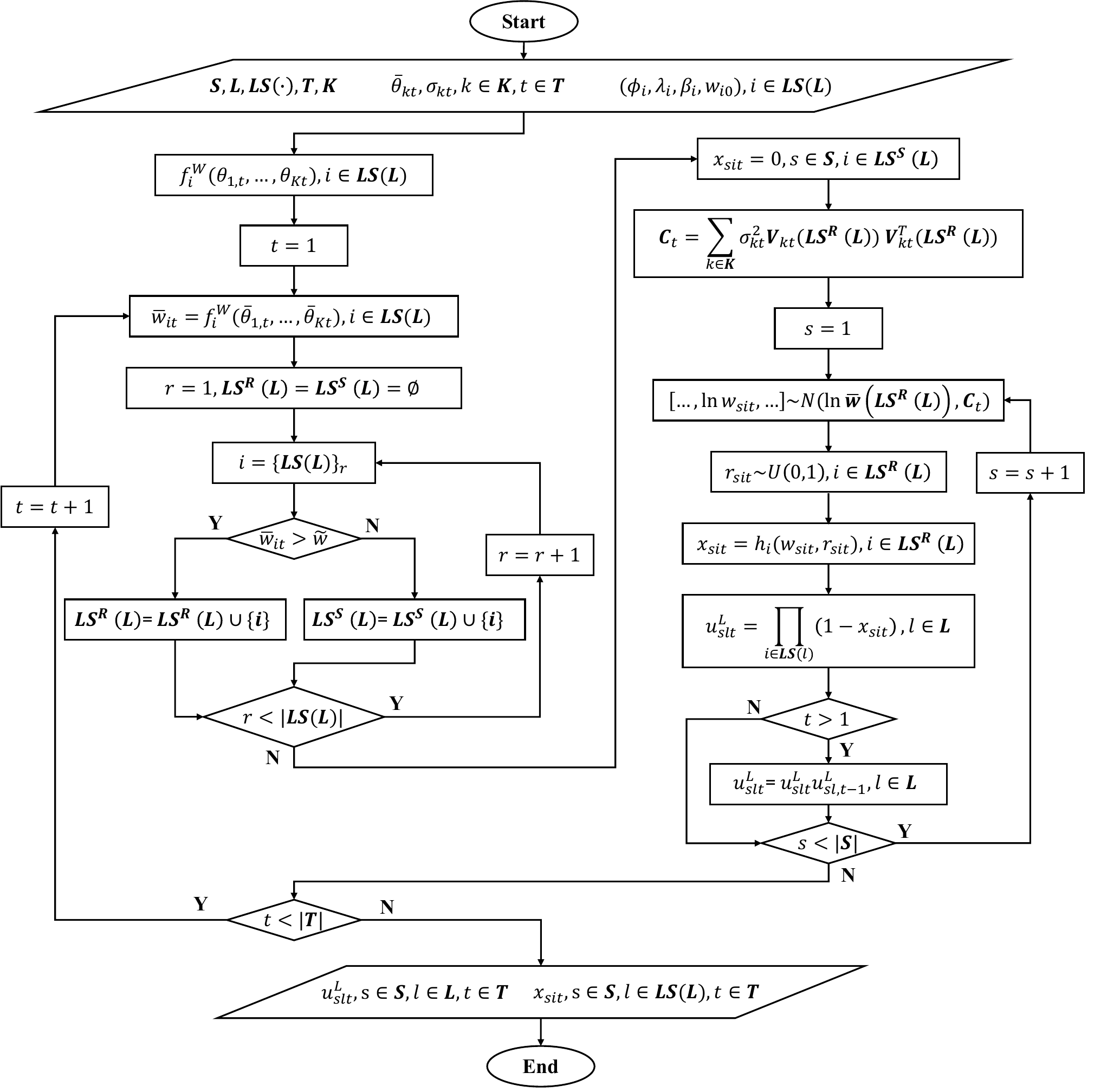}}
	\caption{Process of spatially dependent sampling}
	\label{fig:relevance_sampling_process}
\end{figure}

The spatially dependent sampling process is demonstrated in Figure~\ref{fig:relevance_sampling_process}. The process begins by inputting the sets of required scenarios $\mathbf{S}$, transmission lines $\mathbf{L}$, segments of each transmission line $\mathbf{LS}(\cdot)$, time intervals $\mathbf{T}$, and hurricane parameters $\mathbf{K}$. It also includes the predicted values of each hurricane parameter $\overline{\theta}_{kt}$ and its standard error $\sigma_{kt}$, as well as the coordinates $(\psi_i, \lambda_i)$ and fragility function parameters $\beta_i, w_{i0}$ for each component (line segment).

Next, a function for each hurricane parameter is constructed for each component's location, and the predicted weather intensity is calculated. Based on this mean weather intensity, all components are classified into a relevance sampling set $\mathbf{LS^R(L)}$ and a non-fragile set $\mathbf{LS^S(L)}$. For all components in the non-fragile set $\mathbf{LS^S(L)}$, they are assumed not to fail.

For all components in the relevance sampling set $\mathbf{LS^R(L)}$, the covariance matrix $\mathbf{C}_t$ is calculated through equation (\ref{eq:Ct_calculation}). Based on this covariance matrix, joint lognormal distributed values $w_{sit}$ and corresponding random variables $r_{sit}$ are generated \raa{(In our study we have independent $r_{sit}$, which are not shared across time interval like HRSRA does)}. 
\raa{A Cholesky decomposition $\mathbf{\mathbf{C}_t} = \mathbf{L}\mathbf{L}^T$ will be made to generate $w_{sit}$. Since we have constructed $\mathbf{C}_t$ as sum of rank-1 matrices. This Cholesky decomposition can be done iteratively with rank-1 updates, reducing complexity from $O(N^3)$ to $O(N^2K)$. $N\times N$ is the size of $\mathbf{C}_t$ and $K$ is the number of uncertain parameters.}
Using $x = h(w, r)$ in equation (\ref{eq:two_component_h_function}), the failure state $x_{sit}$ is calculated. The operation state of each transmission line $u^L_{slt}$ is then determined based on the states of all its segments. A transmission line operates normally only when all its segments do not fail. If any segment fails, the transmission line is considered failed in subsequent time intervals. 
\raa{We make this assumption because the maintenance of high-voltage transmission lines requires aerial work, which can only be carried out after the hurricane has passed. Typically, the repair time is longer than the time window we have for preventive control.}

Finally, the state of each component and transmission line is output for each scenario.

\raa{
The SDS method is not only about generating correlated random variables. Beyond that, it explains why correlation exists (i.e. predicition error of extreme event) and quantifies how strong the correlation is.
}

\section{Preventive Control Method}
\label{sec:preventive_control_method}

In this section, we discuss the general process of scenario-based preventive control, \raa{scenario selection method with severity measurement} and formulate a preventive control stochastic unit commitment model.

\subsection{Process of Preventive Control}

The process of preventive control starts with the generation of a scenario pool $\mathbf{S^P}$, from which a subset $\mathbf{S}$ is selected for the preventive control unit commitment model (which we will formulate in \ref{sec:PC_model}). Using this scenario set $\mathbf{S}$ and the unit commitment model, a solution $\mathbf{u^G(S)}$ for unit commitment is obtained. This commitment result $\mathbf{u^G(S)}$ serves as the preventive control measure against the upcoming hazard.

\subsection{Scenario Selection Method and Severity Measurement}

\raa{Using either SDS or SMC, we can generate a large set of plausible component-failure scenarios. However, due to the computational burden, only a small fraction can be included in the preventive control decision model. In this subsection, we discuss scenario-selection methods.}

\raa{To compare scenarios, we define the severity $q(s)$ of a scenario $s$ as follows. Assume $s$ is the realized scenario and that it is known perfectly. We then solve the preventive control model under scenario $s$. The optimal objective value of this model is $q(s)$; that is, $q(s)$ represents the best achievable outcome given perfect prediction.}

\raa{Because the total number of generated scenarios is large, computing the exact $q(s)$ for every $s$ is infeasible within the available time. We therefore adopt a proxy index $\hat{q}(s)$ whose calculation is much faster than that of $q(s)$. The ideal property of $\hat{q}$ is order preservation: for any $s_i,s_j$, 
$\hat{q}(s_i)\le \hat{q}(s_j)\Longleftrightarrow q(s_i)\le q(s_j)$. Note that $\hat{q}(s)$ is used solely for ranking and is not an estimator of the numerical value of $q(s)$.}

\raa{Based on $\hat{q}(s)$ (and $q(s)$ when available), a scenario selection rule chooses a subset $\mathbf{S}^\pi \subseteq \mathbf{S}$. Each $s\in\mathbf{S}^\pi$ is assigned a weight $\pi_s\ge 0$ with $\sum_{s\in\mathbf{S}^\pi}\pi_s=1$. We will propose several concrete selection rules in the case study. The weighted subset $\mathbf{S}^\pi$ is then used in the preventive control model.}

\raa{Since the objective function \eqref{UL_ObjectiveFunction} uses $\pi_s$ to weight the scenario-specific objective values, we define the severity of a selected set as
$q(\mathbf{S}^\pi) = \sum_{s\in \mathbf{S}^\pi} \pi_s\, q(s)$.
For SUC purposes, we prefer $\mathbf{S}^\pi$ to be an unbiased representation of the full set $\mathbf{S}$, aiming for
$q(\mathbf{S}^\pi) \approx \mathbb{E}[\,q(s)\,]$.
For RUC purposes, we emphasize worst-case protection and therefore seek $\mathbf{S}^\pi$ that drives
$q(\mathbf{S}^\pi) \approx \max_{s\in \mathbf{S}} q(s)$.}

\raa{These two choices represent extremes of risk preference. In practice, one may wish to adopt an intermediate stance: introducing some risk aversion so that $q(\mathbf{S}^\pi)$ lies between $\mathbb{E}[q(s)]$ and $\max_{s\in\mathbf{S}} q(s)$. The value of $q(\mathbf{S}^\pi)$ thus provides a convenient index of the risk preference implied by a given selection rule, which in turn shapes the preventive control outcome and its performance across realized scenarios of varying severity.}

\subsection{Preventive Control Stochastic Unit Commitment Model}
\label{sec:PC_model}

We formulate a preventive control stochastic unit commitment model aimed at determining the optimal unit commitment. In the first stage, the model establishes the optimal commitments for unit commitment (UC) to minimize the overall operational risk of the system, represented by scenarios. In the second stage, the units are dispatched based on the optimal UC, determining the power output of all units and the load curtailment at all buses for each representative scenario.

The objective function (\ref{UL_ObjectiveFunction}) aims to minimize startup costs, shutdown costs, and the expected value of the second-stage costs:
\begin{align}
\label{UL_ObjectiveFunction}
\min
\sum_{g \in \mathbf{G}}\sum_{t \in \mathbf{T}} \left(C^{GU}_k y^G_{gt} + C^{GD}_k z^G_{gt}\right)
+ \sum_{s \in \mathbf{S}} \pi_s C_s
\end{align}

The second-stage operation cost (\ref{LL_ObjectiveFunction}) includes generation costs, over-generation costs, and load curtailment costs:
\begin{align}\label{LL_ObjectiveFunction}
    C_s =
    \sum_{g \in \mathbf{G}} \sum_{t \in \mathbf{T}} \left[ C^G(p^G_{sgt}) + C^{OG}_{it} p^{OG}_{sgt} \right]
    + \sum_{n \in \mathbf{N}} \sum_{t \in \mathbf{T}} C^{LC}_{it} \Delta p^D_{snt}
\end{align}

The first-stage constraints include:
\begin{align}
& u^G_{gt} - u^G_{k,t-1} = y^G_{gt} - z^G_{gt},\qquad \forall g \in \mathbf{G}, \forall t \in \mathbf{T} \label{UL_StartStopConstraint}\\
& u^G_{gt} \geq \sum\nolimits_{r=1}^{T^{GU}_g} y^G_{g,t-r+1},\qquad \forall g \in \mathbf{G}, \forall t \in \mathbf{T} \label{UL_MinOnTimeConstraint}\\
& u^G_{gt} \geq \sum\nolimits_{r=1}^{T^{GD}_g} z^G_{g,t+r},\qquad \forall g \in \mathbf{G}, \forall t \in \mathbf{T} \label{UL_MinOffTimeConstraint}\\
& u^G_{gt}, y^G_{gt}, z^G_{gt} \in \{0, 1\},\qquad \forall g \in \mathbf{G}, t \in \mathbf{T} \label{UL_BinaryConstraint}
\end{align}

Equation (\ref{UL_StartStopConstraint}) establishes the relationship between the three sets of binary variables related to the startup and shutdown processes of thermal units. Inequalities (\ref{UL_MinOnTimeConstraint}) and (\ref{UL_MinOffTimeConstraint}) represent the minimum up-time and down-time constraints. Equation (\ref{UL_BinaryConstraint}) ensures that all decision variables at this level are binary.

In the second stage, each scenario is described by $u^L_{slt}$. The constraints include:
\begin{align}\label{LL_DCFlowConstraint}
p^L_{slt} = \frac{\theta^N_{F_l,t} - \theta^N_{T_l,t}}{X^L_l}, \quad \forall s \in \mathbf{S}, \forall l \in \mathbf{L}, u^L_{slt} = 1, t \in \mathbf{T}
\end{align}
\begin{align}\label{LL_PowerBalanceConstraint}
\sum_{g \in \mathbf{G}_i}p^G_{sgt}-
\sum_{l\in\mathbf{L},u^L_{slt}=1,F_l=i}p^L_{slt}+\sum_{l\in\mathbf{L},u^L_{slt}=1,T_l=i}p^L_{slt}=
&p^D_{snt}, \notag\\
\forall s \in \mathbf{S}, \forall n \in \mathbf{N},t \in \mathbf{T} &
\end{align}
\begin{align}\label{LL_DemandBalanceConstraint}
p^D_{snt} = P^{D}_{nt} - \Delta p^D_{snt}, \quad \forall s \in \mathbf{S}, \forall n \in \mathbf{N}, t \in \mathbf{T}
\end{align}
\begin{align}\label{LL_PowerOutputConstraint}
\underline{P}^{G}_k - p^{OG}_{sgt} \leq p^G_{sgt} \leq \overline{P}^{G}_k, \quad \forall s \in \mathbf{S}, \forall g \in \mathbf{G}, t \in \mathbf{T}
\end{align}
\begin{align}\label{LL_LinePowerFlowConstraint}
-\overline{P}^{L}_l \leq p^L_{slt} \leq \overline{P}^{L}_l, \quad \forall s \in \mathbf{S}, \forall l \in \mathbf{L}, u^L_{slt} = 1, t \in \mathbf{T}
\end{align}
\begin{align}\label{LL_load_positive}
p^D_{snt}, \Delta p^D_{snt} \geq 0, \quad \forall s \in \mathbf{S}, \forall n \in \mathbf{N}, t \in \mathbf{T}
\end{align}
\begin{align}\label{LL_load_positive2}
p^G_{sgt}, p^{OG}_{sgt} \geq 0, \quad \forall s \in \mathbf{S}, \forall g \in \mathbf{G}, t \in \mathbf{T}
\end{align}
\begin{align}\label{LL_PhaseAngleConstraint}
\underline{\theta}^{N}_{n} \leq \theta^{N}_{snt} \leq \overline{\theta}^{N}_{n}, \quad \forall s \in \mathbf{S}, \forall n \in \mathbf{N}, t \in \mathbf{T}
\end{align}
\begin{align}\label{LL_RampingConstraint}
\Delta P^{RD}_g \leq p^G_{sgt} - p^G_{sg,t-1} \leq \Delta P^{RU}_k, \quad \forall s \in \mathbf{S}, \forall g \in \mathbf{G}, t \in \mathbf{T}
\end{align}

Equation (\ref{LL_DCFlowConstraint}) represents the constraints of the DCOPF model, while (\ref{LL_PowerBalanceConstraint}) ensures power balance. Equation (\ref{LL_DemandBalanceConstraint}) indicates that the load at each bus is determined by demand and load curtailment. Constraints (\ref{LL_PowerOutputConstraint}) to (\ref{LL_LinePowerFlowConstraint}) limit the power outputs of units and the power flows of transmission lines. Inequalities (\ref{LL_load_positive}) and (\ref{LL_load_positive2}) ensure that load curtailments and over-generation are non-negative. Equations (\ref{LL_PhaseAngleConstraint}) and (\ref{LL_RampingConstraint}) define limits on phase angles and ramping of units, respectively.

\section{Case Study}
\label{sec:case_study}

\subsection{Case System and Methodology Overview}
\label{sec:case_system_methodology}

This case study is conducted using a synthetic power grid model of Texas, USA, based on the ACTIVSg2000 case developed by Texas A\&M University \cite{ACTIVSg2000}. The model provides a detailed representation of Texas’s power grid using publicly available data. The grid in our study comprises 1551 buses operating at 115kV or higher, interconnected by 2749 transmission lines, and supported by 544 generation units, with a total installed capacity of 100,085 MW.

\begin{figure}
    \centering
    \includegraphics[width=0.8\linewidth]{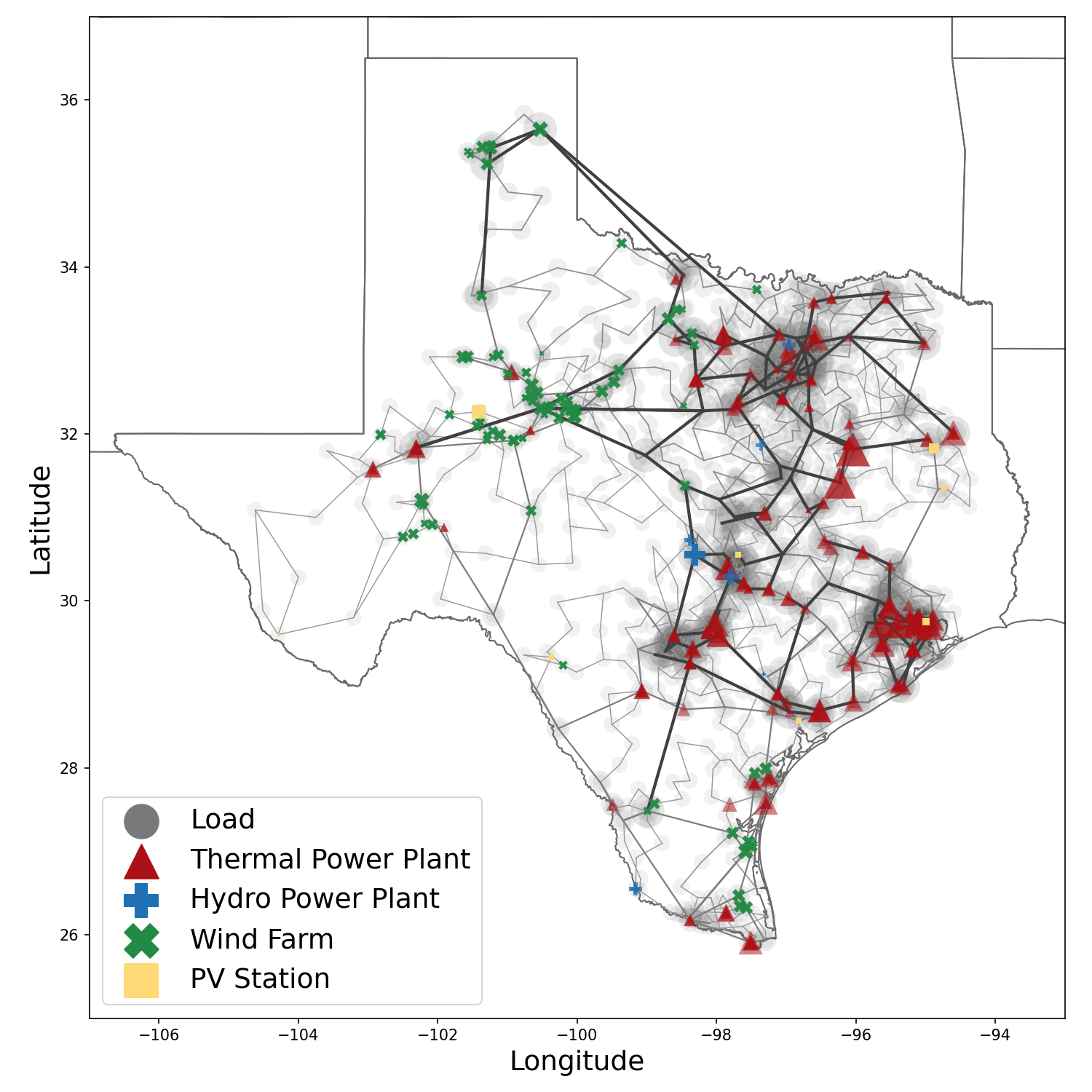}
    \caption{Overview of the Case System}
    \label{fig_case_system_overview}
\end{figure}

The study covers a 24-hour period from August 25 to 26, 2017, coinciding with the landfall of Hurricane \textit{Harvey} on the southeastern coast of Texas. During this period, the total system load fluctuated between a low of 55,271 MW and a high of 81,362 MW (scaled from the original load profile).

\begin{figure}
    \centering
    \subfigure[t=1]{
        \begin{minipage}[t]{0.4\linewidth}
            \centering
            \includegraphics[width=1\linewidth]{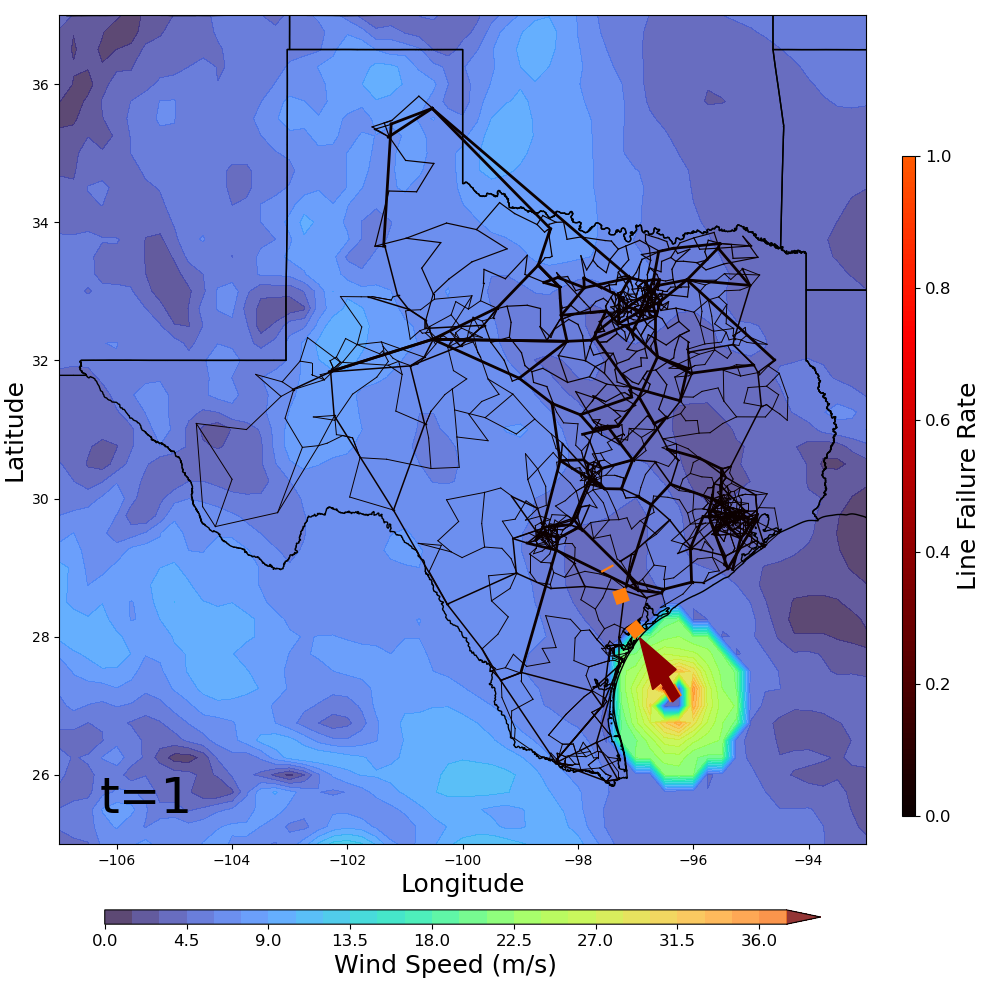}
        \end{minipage}%
    }
    \subfigure[t=8]{
        \begin{minipage}[t]{0.4\linewidth}
            \centering
            \includegraphics[width=1\linewidth]{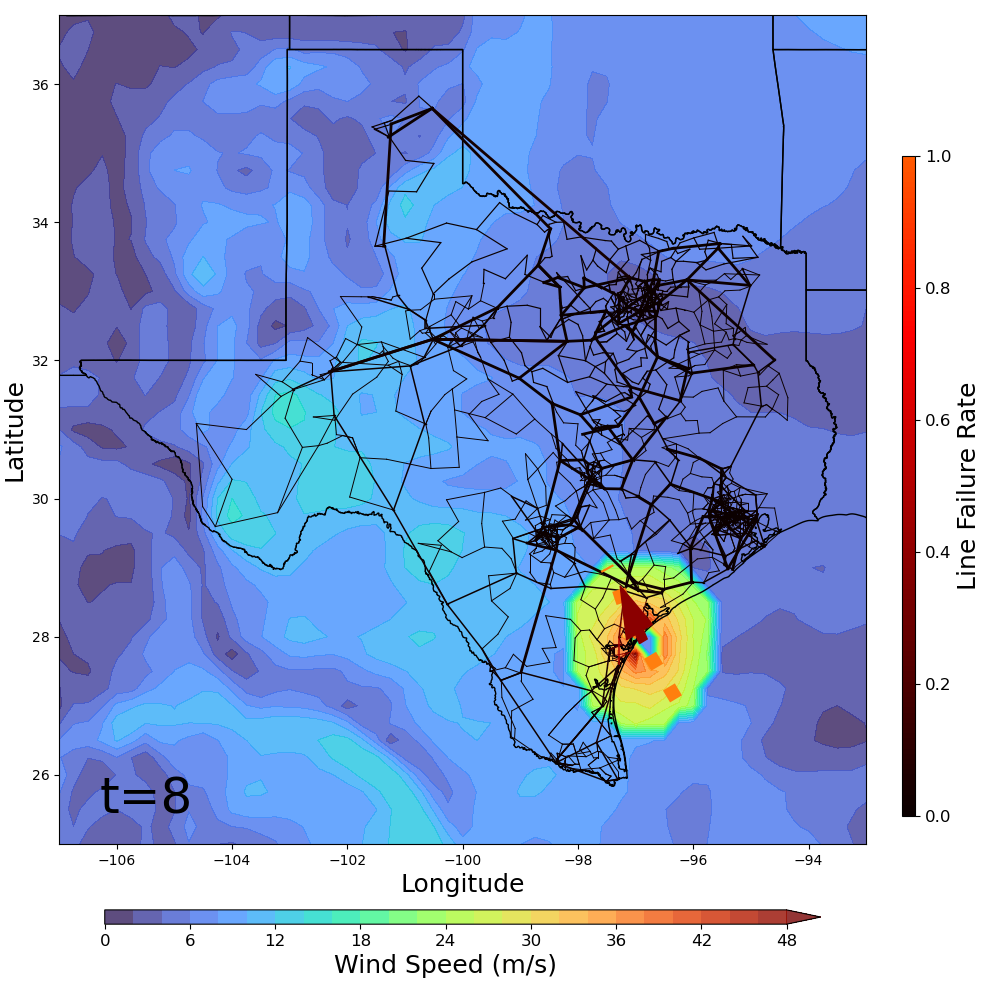}
        \end{minipage}%
    }%
    \\
    \subfigure[t=16]{
        \begin{minipage}[t]{0.4\linewidth}
            \centering
            \includegraphics[width=1\linewidth]{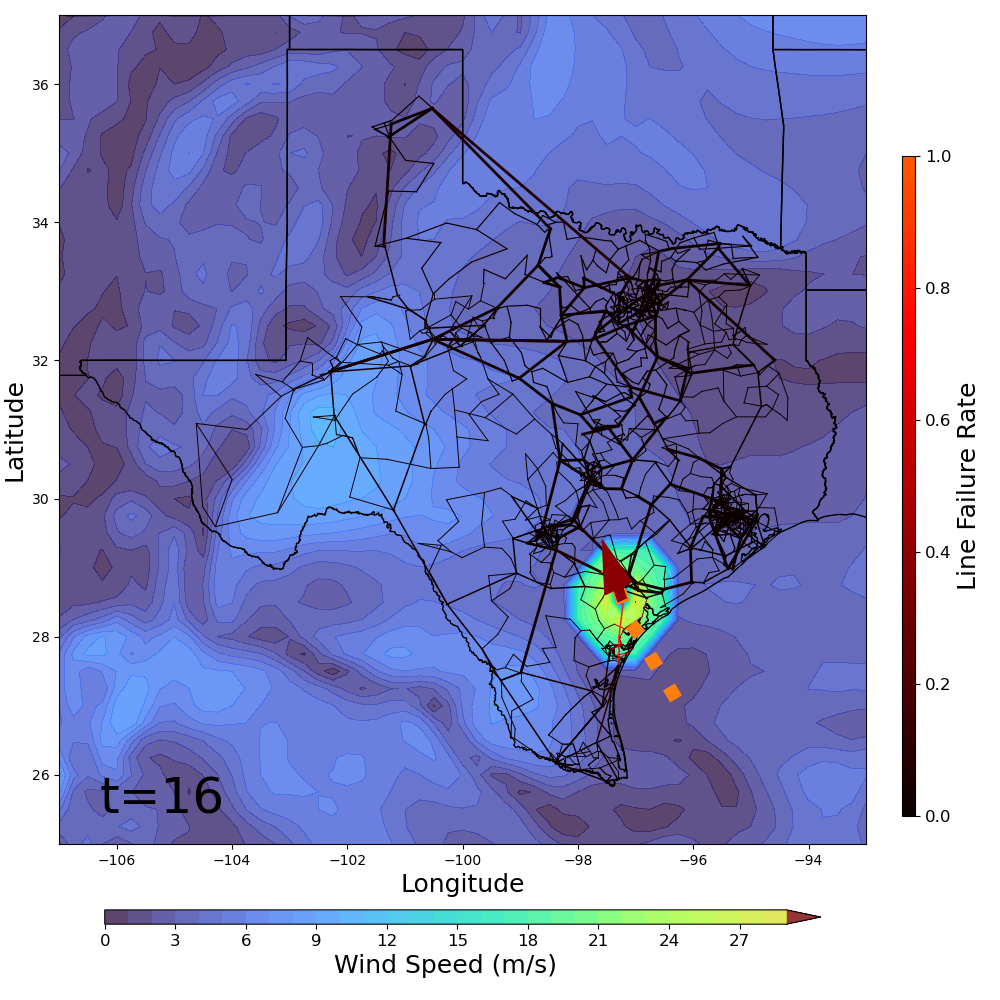}
        \end{minipage}%
    }
    \subfigure[t=24]{
        \begin{minipage}[t]{0.4\linewidth}
            \centering
            \includegraphics[width=1\linewidth]{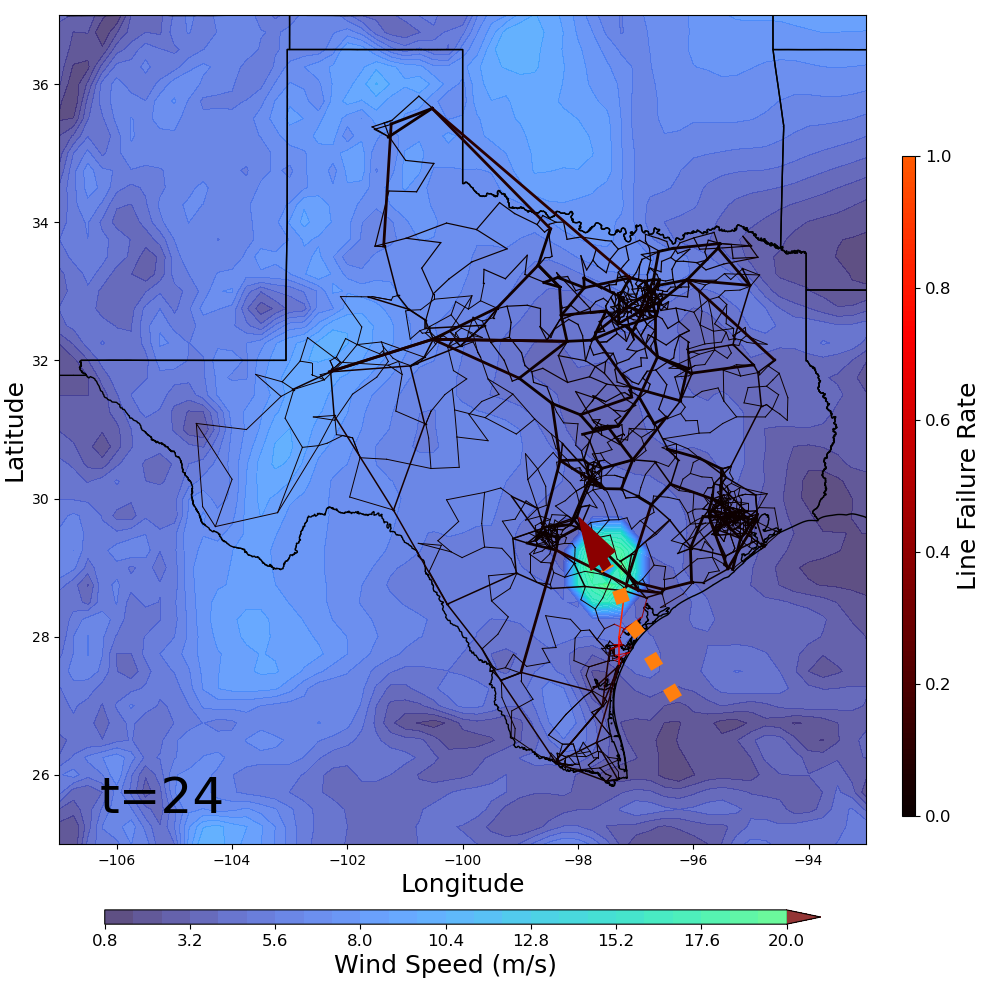}
        \end{minipage}%
    }
    \caption{Path, wind speed, and range of the synthetic hurricane}
    \label{fig_hurricane_path}
\end{figure}

The simulated hurricane's impact on the southeastern Texas coast mimics the actual trajectory and extent of Hurricane \textit{Harvey}, as shown in Figure~\ref{fig_hurricane_path}. 
\raa{We use a Holland hurricane wind speed field model in our study, as described in \ref{app:holland}. And we directly use the wind speed calculated from the model as weather intensity.}
Key moments include:
\begin{itemize}
    \item At t=1, the hurricane makes landfall on the southeastern coast of Texas.
    \item At t=8, the hurricane's eye reaches Texas, causing the most significant damage.
    \item At t=16, the hurricane moves northeast across Texas, with decreasing size and wind speed.
    \item By t=24, the hurricane weakens significantly and exits Texas.
\end{itemize}

The methods employed in this study include scenario generation, scenario selection, and dispatch techniques. The simplified names and detailed explanations of the scenario generation methods are provided below.
\begin{itemize}
    \item \textbf{Relevance}: The spatially dependent sampling method, denoted as relevance sampling for short.
    \item \textbf{Normal}: The commonly used sequential Monte Carlo simulation method, denoted as normal sampling for short.
\end{itemize}

\begin{figure}
	\centering
	\subfigure[t=1]{
		\begin{minipage}[t]{0.4\linewidth}
			\centering
			\includegraphics[width=1\textwidth]{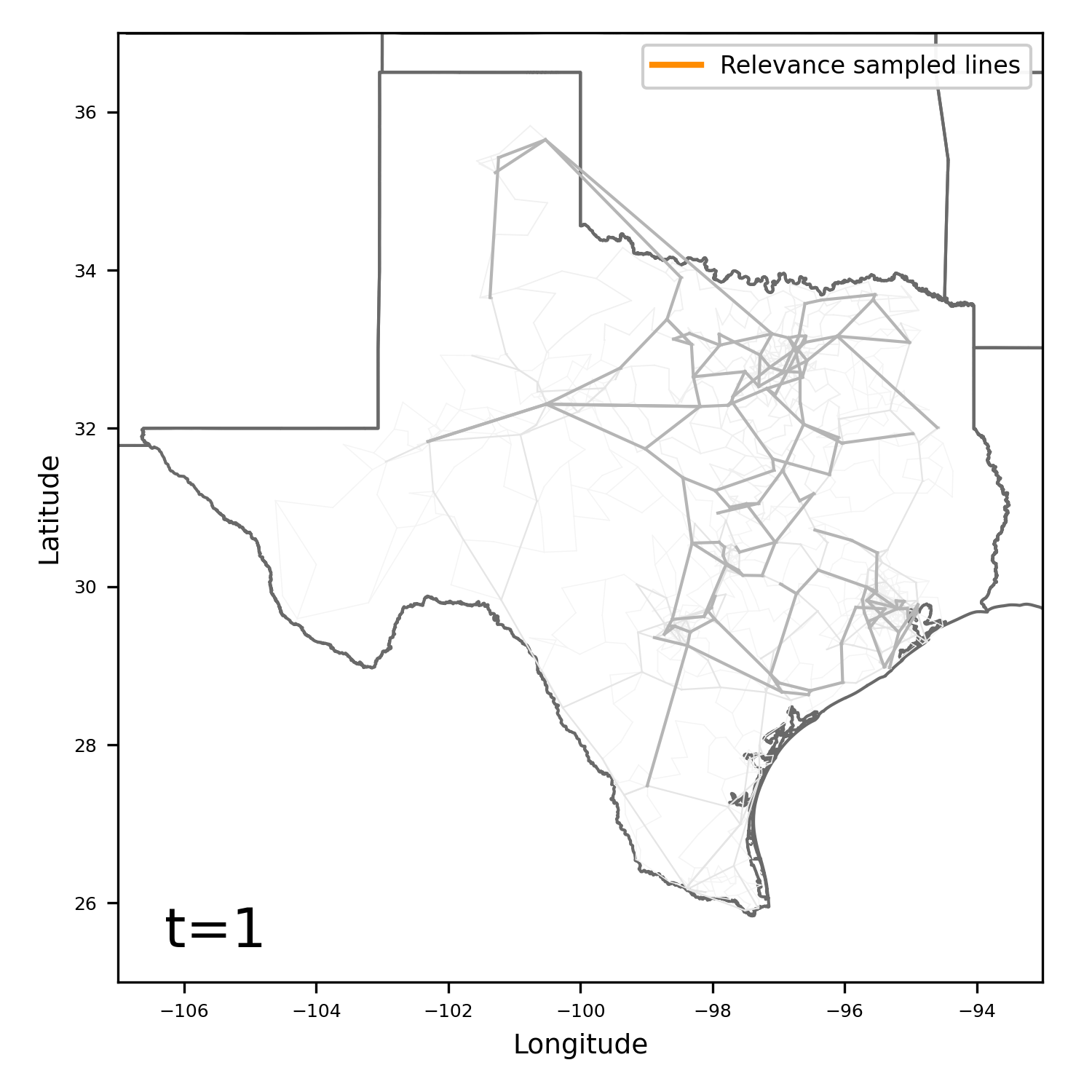}
		\end{minipage}
	}
	\subfigure[t=8]{
		\begin{minipage}[t]{0.4\linewidth}
			\centering
			\includegraphics[width=1\textwidth]{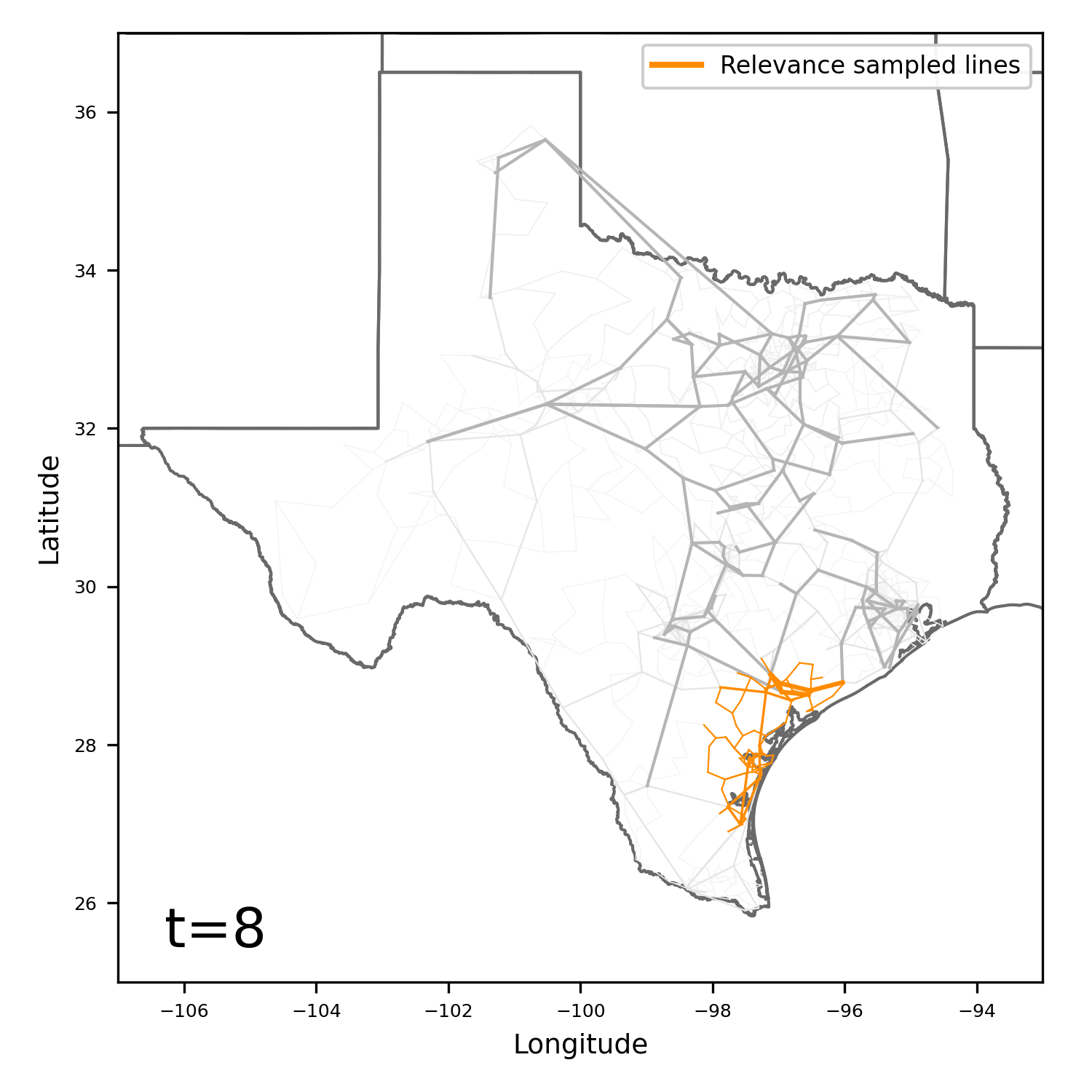}
		\end{minipage}
	}
	\\
	\subfigure[t=16]{
		\begin{minipage}[t]{0.4\linewidth}
			\centering
			\includegraphics[width=1\textwidth]{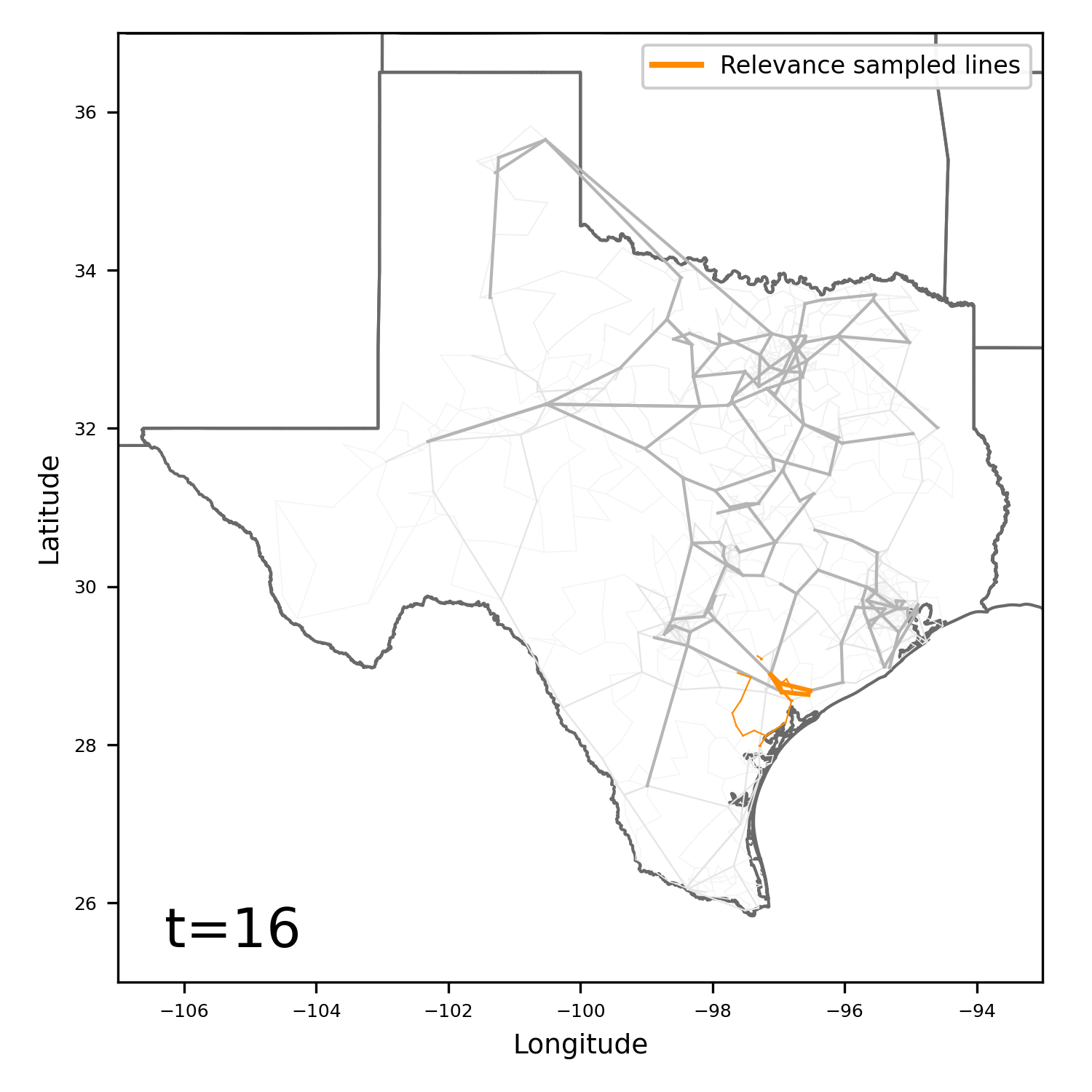}
		\end{minipage}
	}
	\subfigure[t=24]{
		\begin{minipage}[t]{0.4\linewidth}
			\centering
			\includegraphics[width=1\textwidth]{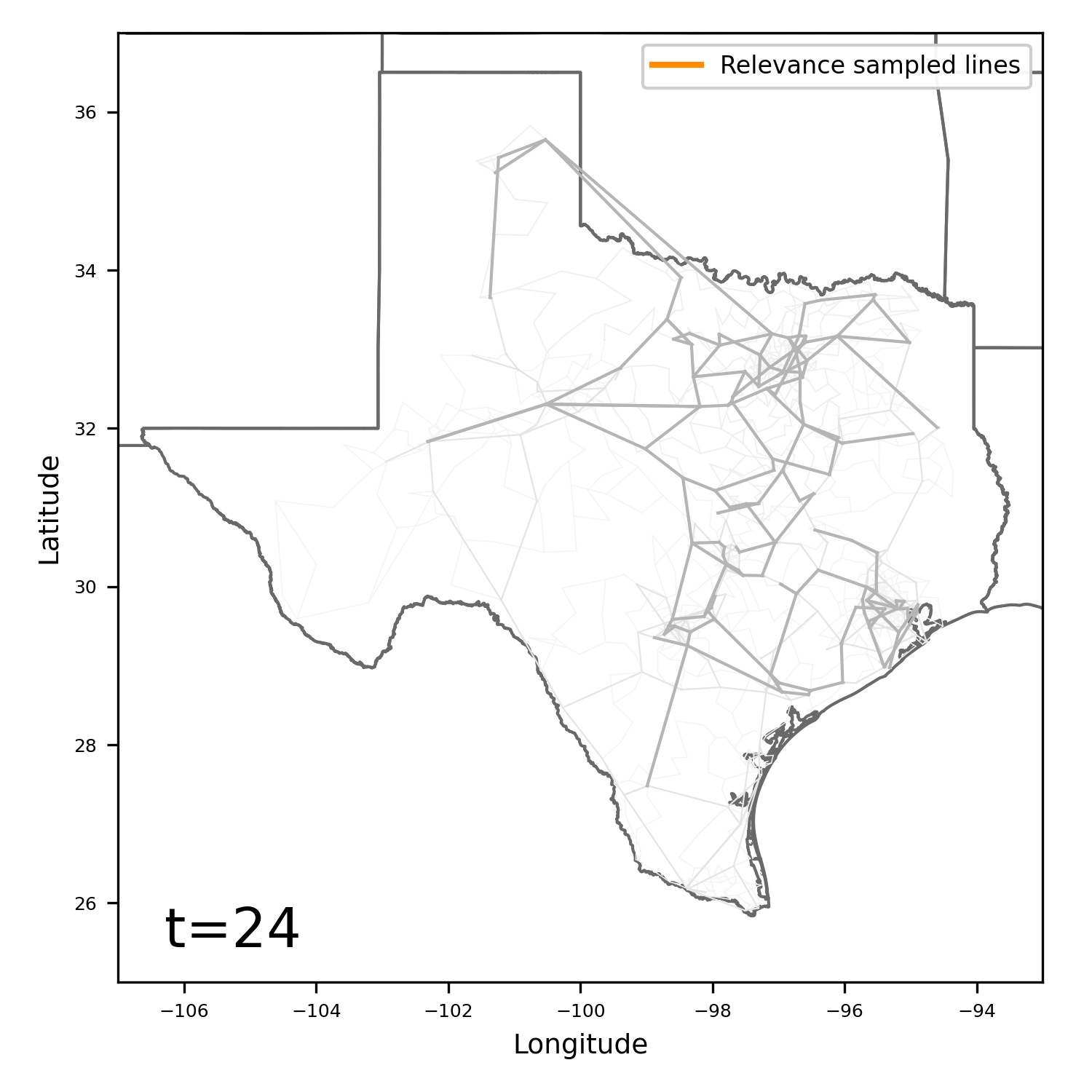}
		\end{minipage}
	}
	\caption{Transmission lines sampled by the relevance method}
	\label{fig_rel_sampled_lines}
\end{figure}

This study investigates the failure of transmission lines during a hurricane event. To improve granularity, the lines are divided into smaller segments. Lines selected by the relevance sampling method at specific time intervals are highlighted in Figure~\ref{fig_rel_sampled_lines}. The number of relevance-sampled lines increases as the hurricane's intensity intensifies.

The scenario selection methods are as follows:
\begin{itemize}
    \item \textbf{N-random}: N scenarios randomly selected from the total set of possible scenarios. \raa{Each scenario is assigned an equal weight $\pi_s=1/N$.}
    \item \textbf{N-stratified}: N scenarios selected from all scenarios through stratified sampling according to the number of failed transmission lines. \raa{Weight $\pi_s$ is equal within strata and strata weight is proportional to pool mass.}
    \item \textbf{N-worst}: Top N scenarios with the highest number of failed transmission lines. \raa{Each scenario is assigned an equal weight $\pi_s=1/N$.}
\end{itemize}

\raa{
These three scenario sets (random, stratified and worst) can select scenarios with different level of severity. They will be used to demonstrate how varying risk preferences influence the outcomes of the preventive control model.
}

The dispatch methods are described as:
\begin{itemize}
    \item \textbf{RES- x\%}: Deterministic unit commitment with no faults, considering an x\% reserve.
    \item \textbf{PC N-xxx}: Preventive control unit commitment model considering scenarios selected by the \textbf{N-xxx} method.
\end{itemize}

\begin{figure}
    \centering
    \includegraphics[width=0.9\linewidth]{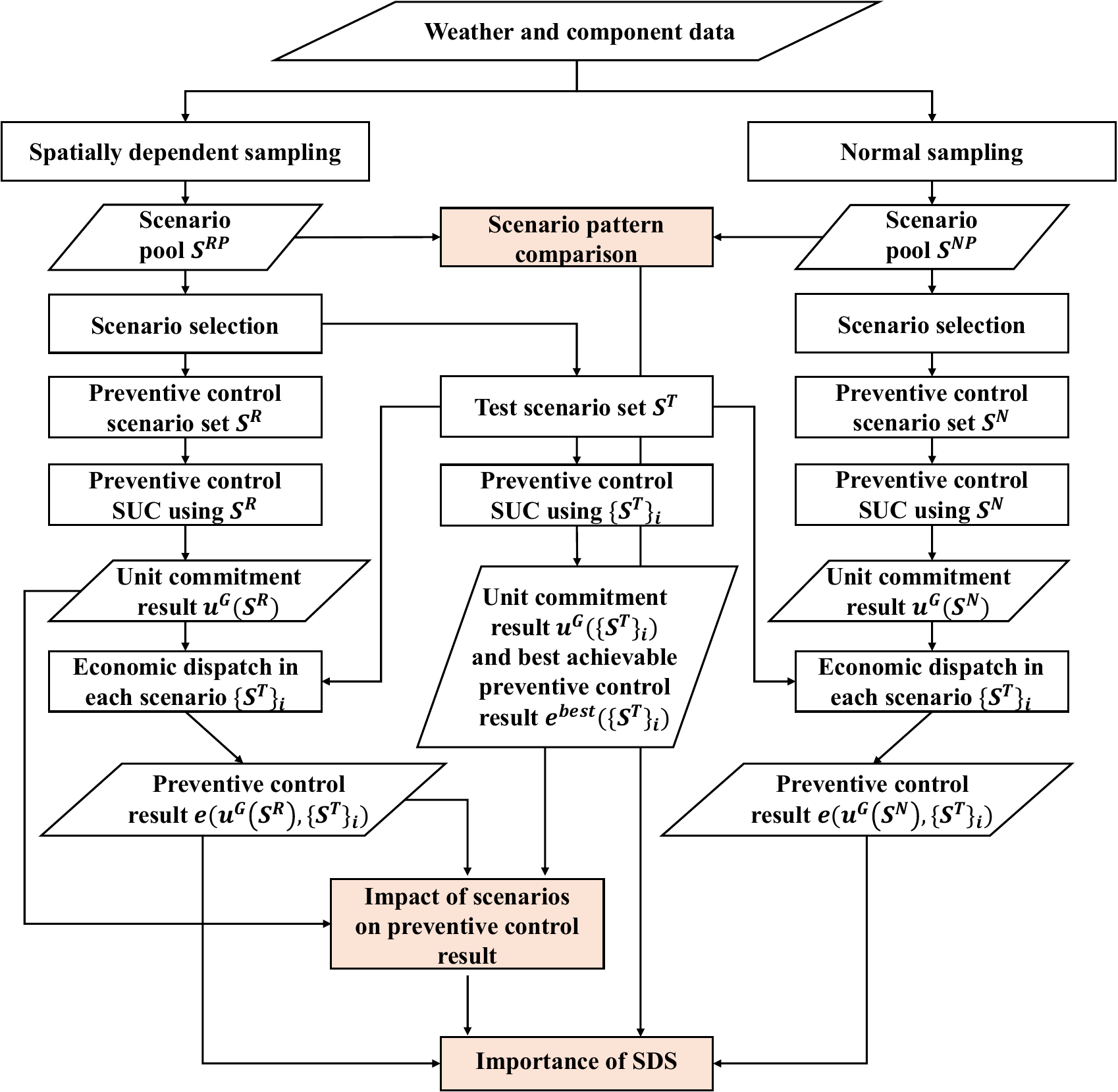}
    \caption{\raa{Comparison framework for the case study section}}
    \label{fig:case_study_comparison_framework}
\end{figure}

\raa{
The methodology framework for this case study is presented in Figure~\ref{fig:case_study_comparison_framework}. Using identical weather and component data, we generate scenario pools $\mathbf{S^{RP}}$ and $\mathbf{S^{NP}}$ through relevance and normal sampling. The patterns of $\mathbf{S^{RP}}$ and $\mathbf{S^{NP}}$ are compared in Section \ref{sec:scenario_pattern}. Subsets of scenarios, $\mathbf{S^{R}}$, $\mathbf{S^{N}}$, and $\mathbf{S^{T}}$, are selected based on the scenario selection methods and implemented in the preventive control model to obtain unit commitment results $\mathbf{u^G}(\mathbf{S^R})$. These unit commitment results are then used to conduct economic dispatch in each realized scenario from $\mathbf{S^T}$, yielding dispatch results $\mathbf{e}(\mathbf{u^G}(\mathbf{S^R}),\{\mathbf{S^T}\}_i)$. The dispatch outcomes are compared in Section \ref{sec:impact_of_scenario_selection} to investigate the impact of scenario severity on preventive control results. Further comparisons of $\mathbf{u^G}(\mathbf{S^R})$ and $\mathbf{u^G}(\mathbf{S^N})$ aim to identify the underlying reasons for any differences observed. Finally, in Section \ref{sec:importance_of_SDS} we obtain $\mathbf{u^G}(\mathbf{S^N})$ similarly and test them on $\mathbf{S^T}$ to obtain $\mathbf{e}(\mathbf{u^G}(\mathbf{S^N}),\{\mathbf{S^T}\}_i)$. We compare $\mathbf{e}(\mathbf{u^G}(\mathbf{S^R}),\{\mathbf{S^T}\}_i)$ and $\mathbf{e}(\mathbf{u^G}(\mathbf{S^N}),\{\mathbf{S^T}\}_i)$ to show the importance of SDS.
}

\raa{For the scalability issue, here we report several key value. We conduct the experiments on a workstation with AMD Ryzen 9950X CPU and 192GB RAM, and all codes are implemented with Python 3.10. Since we set a wind speed threshold for component failure sampling, the size of $\mathbf{C}_t$ is approximately $8000\times 8000$ when hurricane is strong, and drop to below $500\times 500$ when hurricane is weak. For one large $\mathbf{C}_t$, the construction time is approximately 1s, and for one small $\mathbf{C}_t$ the construction time is below 0.05s. The end-to-end runtime of drawing 10,000 scenarios is approximately 100s. The only difference between SDS and normal SMC method is that SDS has an additional process of building $\mathbf{C}_t$, Cholesky decomposition $\mathbf{\mathbf{C}_t} = \mathbf{L}\mathbf{L}^T$ and independent random variables transformation with $\mathbf{L}$ for each time interval. The runtime of this process is less than 25s for the 24h we have studied, which is the additional runtime of SDS compared with SMC. The solution time of solving a business-as-usual UC (i.e. considering no failure) is around 100s. The solution time of solving an SUC with 5 and 10 failure scenarios are approximately 1500s and 4000s respectively.
}

\subsection{Scenario Pattern Comparison}
\label{sec:scenario_pattern}

\begin{figure}
	\centering
	\includegraphics[width=0.8\linewidth]{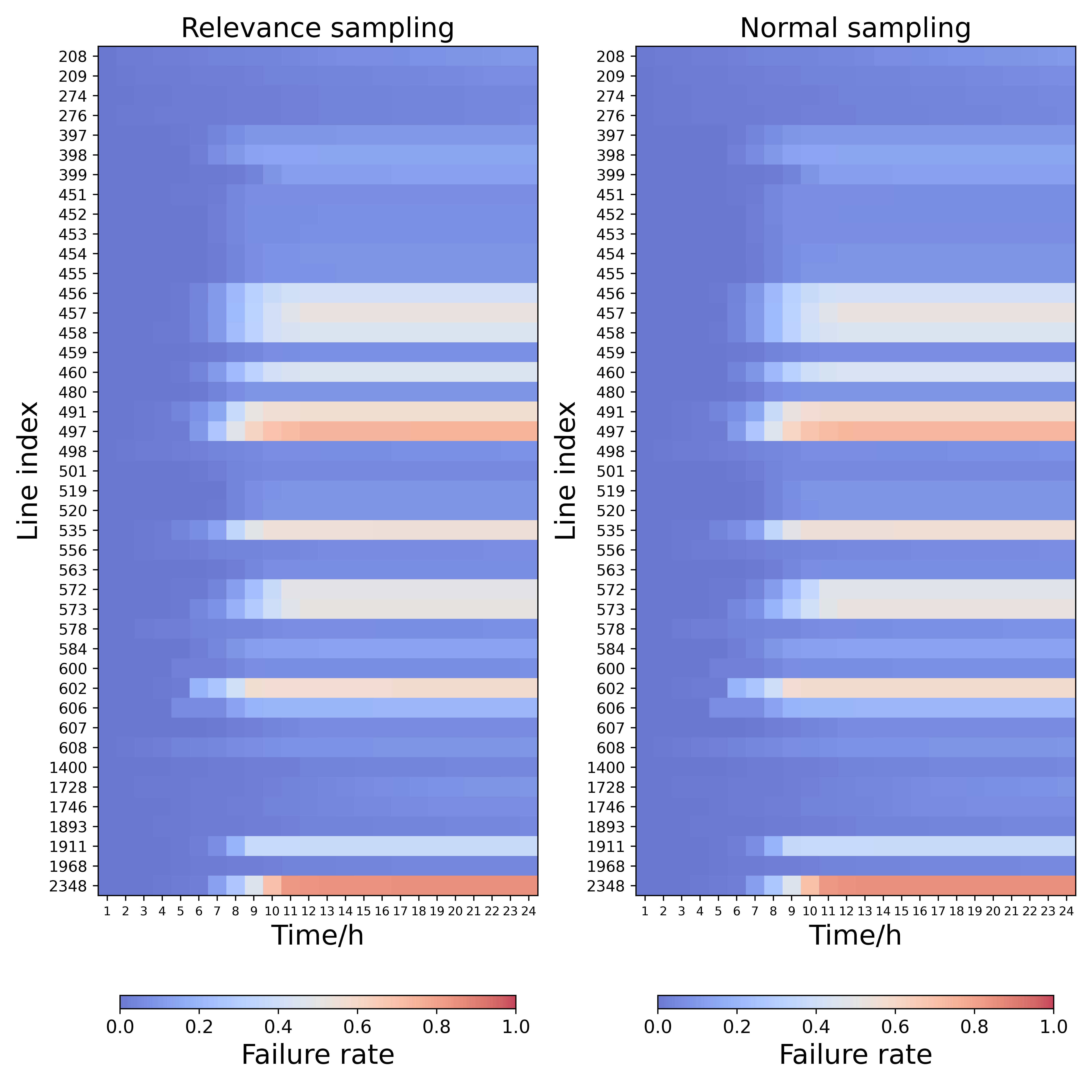}
	\caption{Failure probability of each line across time intervals}
	\label{fig_line_FO_prob}
\end{figure}

We applied both relevance sampling and normal sampling techniques to generate 10,000 scenarios respectively. The failure probabilities of individual transmission lines at different time intervals, based on both scenario sets, are illustrated in Figure~\ref{fig_line_FO_prob}. The results show that the failure probabilities are nearly identical for both methods, indicating they yield comparable expected outcomes. Notably, expected failure counts are often used to calibrate fragility parameters.

\begin{figure}
	\centering
	\includegraphics[width=0.7\linewidth]{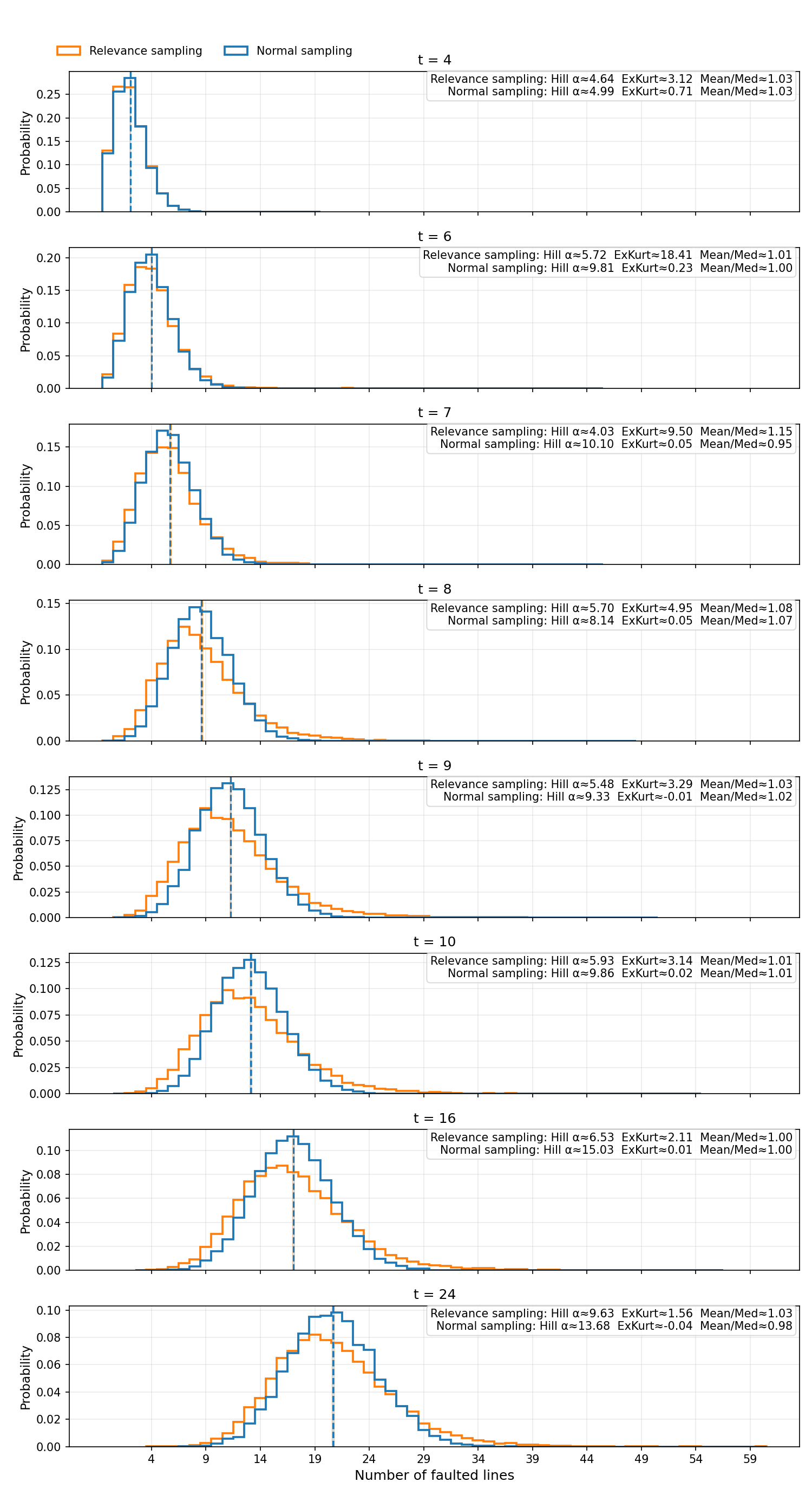}
	\caption{\raa{Distribution of faulted line number}}
	\label{fig_line_FO_dist}
\end{figure}

Although the overall failure probabilities appear similar between the two sampling methods, the distributions of these failures differ substantially, as shown in Figure~\ref{fig_line_FO_dist}. 
\raa{Distributions of the number of faulted lines at each time interval for scenarios generated by relevance sampling and normal sampling are shown in orange and blue, respectively. A vertical dashed line marks the mean of each distribution. Over the displayed interval, both methods yield nearly identical mean counts of faulted lines.}

\raa{
We evaluate heavy-tailed behavior in the distribution of faulted line counts using three complementary metrics: the Hill tail index, excess kurtosis, and the mean-to-median ratio (MMR). The metrics, their abbreviations as shown in Figure~\ref{fig_line_FO_dist}, and the interpretation of their magnitudes with respect to tail heaviness are summarized in Table~\ref{tab:tail-metrics}.
}

\begin{table}[h]
\centering
\caption{\raa{Tail-heaviness metrics used in this paper and their interpretation.}}
\label{tab:tail-metrics}
\begin{tabularx}{\linewidth}{@{}l c X@{}}
\toprule
\textbf{Metric} & \textbf{Plot abbreviation} & \textbf{Value magnitude and heavy-tail relationship} \\
\midrule
Hill tail index $\alpha$ & \textbf{Hill $\alpha$} &
Smaller $\alpha$ $\Rightarrow$ heavier right tail. \\

Excess kurtosis $\kappa$ & \textbf{ExKurt} &
Larger $\kappa$ $\Rightarrow$ more peaked and heavier tails (Gaussian baseline $\kappa=0$).\\

Mean-to-median ratio & \textbf{MMR} &
$\mathrm{MMR}\approx 1$ for symmetric or thin-tailed distributions; larger $\mathrm{MMR}>1$ indicates a heavier right tail pulling up the mean. \\
\bottomrule
\end{tabularx}
\end{table}

\raa{
Relevance sampling yields heavier tails in the distribution of transmission line failures. The relevance-sampled distributions consistently exhibit smaller Hill tail indices $\alpha$, larger excess kurtosis $\kappa$, and mean-to-median ratios greater than 1, all indicative of heavy-tailed behavior. Consequently, relevance sampling captures more extreme scenarios with a larger number of simultaneous line failures. 
}

\raa{Although these more extreme scenarios may not be actually realized, the results suggest that normal sampling underestimates their risk. By overlooking plausible extremes, it invites catastrophic outcomes when such conditions do occur and the system is unprepared.}

\subsection{Impact of Scenarios on Preventive Control Results}
\label{sec:impact_of_scenario_selection}

We conduct a comparative analysis involving two types of approaches: (1) scenario-based preventive control strategies that consider different sets of potential contingency scenarios, and (2) deterministic methods that do not account for potential failure scenarios.
By comparing scenario-based preventive control with deterministic methods, we examine the impact of incorporating anticipated scenarios on system performance.
By comparing preventive control strategies based on different scenario sets, we investigate how the choice of scenario sets influences the results.

Each method produces a unit commitment plan, which is then used in economic dispatch across 100 test scenarios $\mathbf{S^T}$, randomly selected from a relevance-based pool $\mathbf{S^{RP}}$ of 10,000 scenarios. 
\raa{
We perform stratified sampling of all scenarios by fault count, using deciles. The scenarios are partitioned into ten strata, and we sample ten scenarios from each stratum. The selected 100 test scenarios, denoted $\mathbf{S}^T$, provide a benchmark for evaluating each strategy. For every $s\in\mathbf{S}^T$, we solve for the optimal preventive-control plan $\mathbf{e}^{\text{best}}(s)$ and record its severity $q(s)$.
}
\raa{
In Section~\ref{sec:impact_of_scenario_selection}, $\mathbf{S}^T$ is a testbed that spans low-to-high severities for comparing selection rules and their implied risk preferences. It is not assumed to be a ground truth.
}

\begin{figure}
	\centering
	\includegraphics[width=0.75\linewidth]{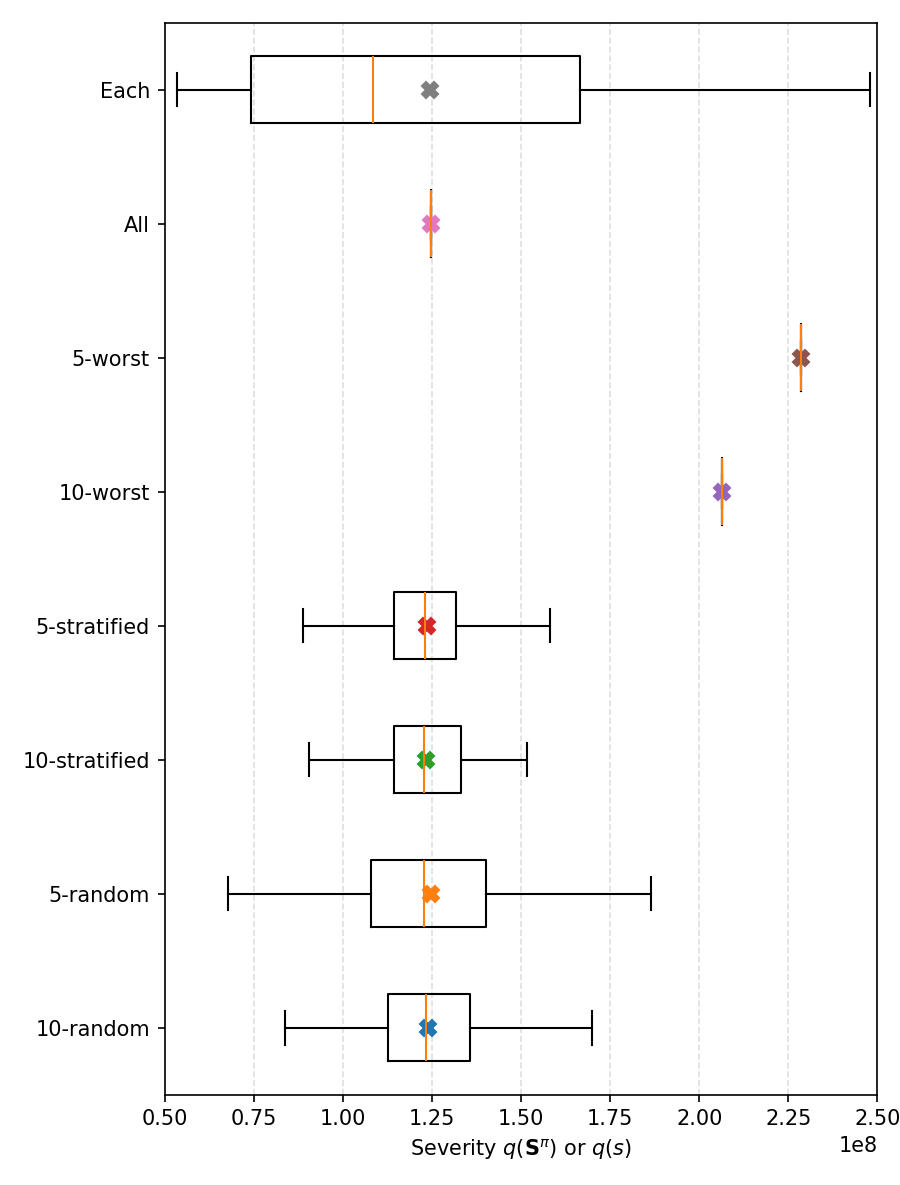}
	\caption{\raa{Severity distribution of $\mathbf{S}^\pi \subseteq \mathbf{S^T}$ selected by each method}}
	\label{fig:severity_dist_ST}
\end{figure}

\raa{
We consider three selection rules (\emph{$N$-random}, \emph{$N$-stratified} and \emph{$N$-worst}) and form subsets $\mathbf{S}^\pi$ of size $5$ and $10$ (six rules in total). Each rule selects scenarios from $\mathbf{S}^{RP}$. Because $q(s)$ has been computed for all $s\in\mathbf{S}^T$, we also use $\mathbf{S}^T$ as a testbed to assess the selection rules. For the \emph{$N$-random} and \emph{$N$-stratified} rules, the set severity $q(\mathbf{S}^\pi)$ is a random variable; we display their distributions via box plots in Figure~\ref{fig:severity_dist_ST}. A vertical line marks the median, and a ``$\times$'' marks the mean. For reference, the row ``Each'' shows the distribution of individual severities $q(s)$ over $\mathbf{S}^T$, and the row ``All'' corresponds to the rule that retains all input scenarios (i.e., a perfect unbiased selection for stochastic unit commitment).
}

\raa{
For both the \emph{$N$-random} and \emph{$N$-stratified} rules, the mean of $q(\mathbf{S}^\pi)$ matches the average severity over all scenarios, indicating that these rules are \emph{unbiased}. The stratified rule yields a tighter distribution than the random rule. In contrast, the \emph{10-worst} and \emph{5-worst} selections produce larger $q(\mathbf{S}^\pi)$ values, approaching the maximum observed severity; thus $N$-worst rules are \emph{biased} toward high-severity scenarios (smaller $N$ implies higher severity). These differences reflect varying degrees of risk preference/aversion. Later we will evaluate how $q(\mathbf{S}^\pi)$ influences the preventive control solution and its performance across realized scenarios.
}

The average dispatch cost is regarded as the expected performance of each \raa{dispatch} method. Table~\ref{tab:cost_cmp_nDR} presents the breakdown of costs: LC (load curtailment), SU/SD (start-up/shut-down), OP (operating), and OG (over-generation).

\begin{table}[t]
	\centering
	\caption{Expected cost of different methods}
    \scalebox{0.6}{
	\begin{tabular}{cccccc}
		\hline
		\specialrule{0em}{1pt}{1pt}
		Method & Expected Total Cost & Expected LC Cost & Expected SU/SD Cost & Expected OP Cost & Expected OG Cost \\
		\specialrule{0em}{1pt}{1pt}
		\hline
		RES-0\% & 12167.28 (105.00\%) & 7988.82 (110.15\%) & 469.46 (92.58\%) & 3269.54 (97.07\%) & 439.45 (95.68\%) \\
		RES-10\% & 11587.37 (100.00\%) & 7252.77 (100.00\%) & 507.07 (100.00\%) & 3368.24 (100.00\%) & 459.28 (100.00\%) \\
		RES-20\% & 11318.05 (97.68\%) & 6905.49 (95.21\%) & 493.88 (97.4\%) & 3472.42 (103.09\%) & 446.26 (97.16\%) \\
		PC5-random & 10697.49 (92.32\%) & 6504.02 (89.68\%) & 509.51 (100.48\%) & 3378.46 (100.3\%) & 305.49 (66.51\%) \\
		PC5-stratified & 10682.2 (92.19\%) & 6433.26 (88.7\%) & 507.96 (100.17\%) & 3371.33 (100.09\%) & 369.65 (80.48\%) \\
		PC5-worst & 13671.14 (117.98\%) & 9685.2 (133.54\%) & 511.47 (100.87\%) & 3373.9 (100.17\%) & 100.57 (21.9\%) \\
		PC10-random & 10653.58 (91.94\%) & 6431.0 (88.67\%) & 507.92 (100.17\%) & 3379.87 (100.35\%) & 334.8 (72.9\%) \\
		PC10-stratified & 10716.8 (92.49\%) & 6425.69 (88.6\%) & 508.3 (100.24\%) & 3366.72 (99.95\%) & 416.09 (90.6\%) \\
		PC10-worst & 13681.2 (118.07\%) & 9690.96 (133.62\%) & 511.36 (100.85\%) & 3379.78 (100.34\%) & 99.09 (21.58\%) \\
		\specialrule{0em}{1pt}{1pt}
		\hline
	\end{tabular}
    }
	\label{tab:cost_cmp_nDR}
\end{table}

In terms of expected costs, preventive control strategies that incorporate random and stratified scenarios outperform deterministic methods and those focused on worst-case scenarios. Increasing the reserve requirement from 0\% to 20\% significantly reduces the expected total cost, primarily by lowering load curtailment costs. Preventive control methods can reduce expected load curtailment costs by over 10\% and total expected costs by approximately 8\%, highlighting the importance of strategic unit commitment adjustments in anticipation of a hurricane. However, methods focusing on worst-case scenarios result in the highest expected total costs, primarily due to higher load curtailment costs. The findings also suggest that considering five scenarios is sufficient for satisfactory results, as adding more scenarios imposes a computational burden without yielding substantial benefits.

\begin{figure}
    \centering
    \includegraphics[width=0.8\linewidth]{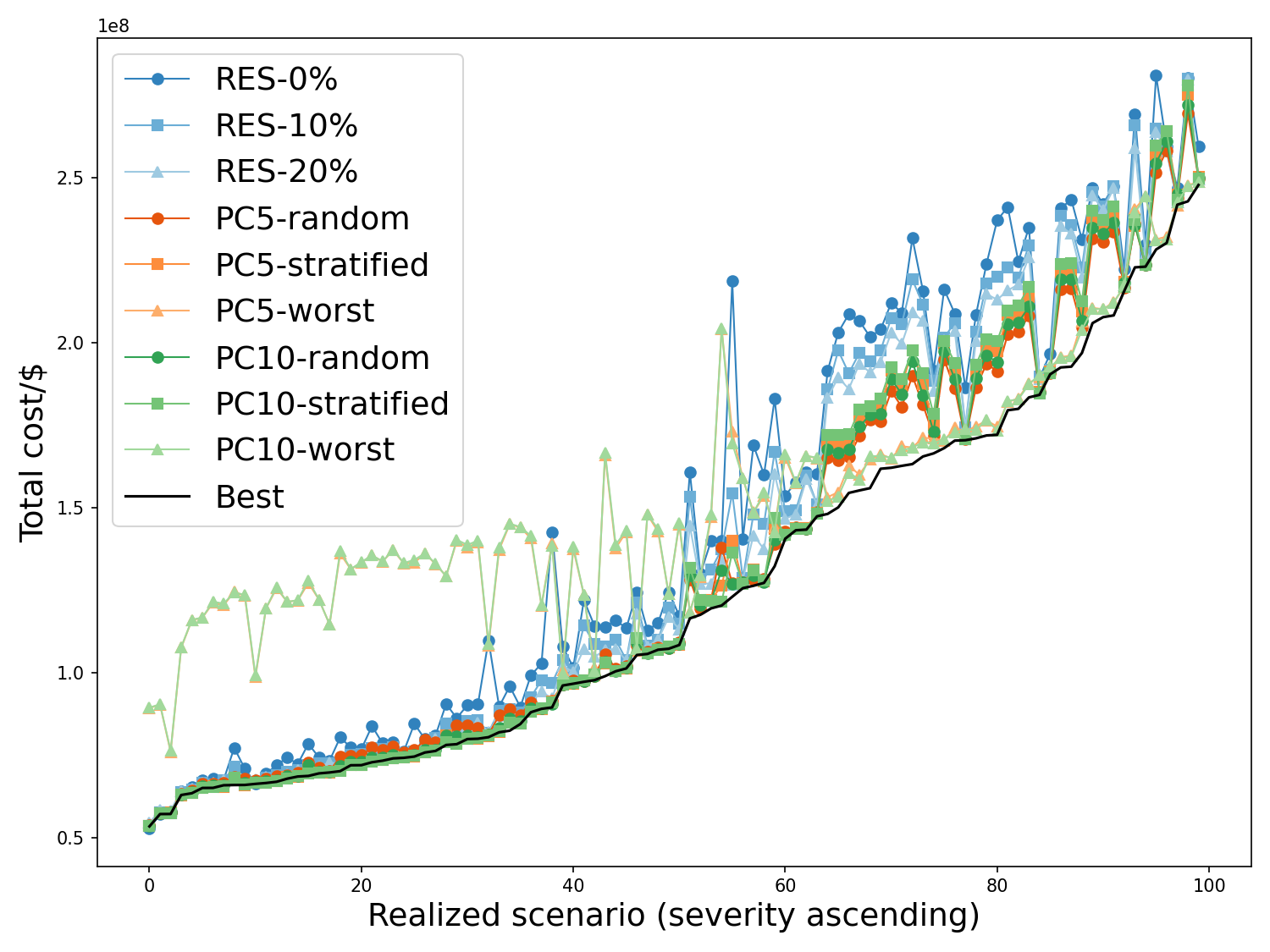}
    \caption{Total cost of different methods in each realized scenario}
    \label{fig_cost_by_sce_obj_nDR}
\end{figure}

Figure~\ref{fig_cost_by_sce_obj_nDR} illustrates the total costs achieved by each method across various scenarios. Preventive control methods considering random and stratified scenarios generally outperform deterministic methods, approaching the optimal result in less severe scenarios. However, as \raa{realized} severity increases, the performance gap widens. 
\raa{Preventive control methods accounting for worst-case scenarios incur higher costs in less severe realized scenarios but converge to the best results in more severe conditions.} 
These findings emphasize the divergent outcomes of stochastic versus robust optimization approaches.

\begin{figure}
    \centering
    \includegraphics[width=0.8\linewidth]{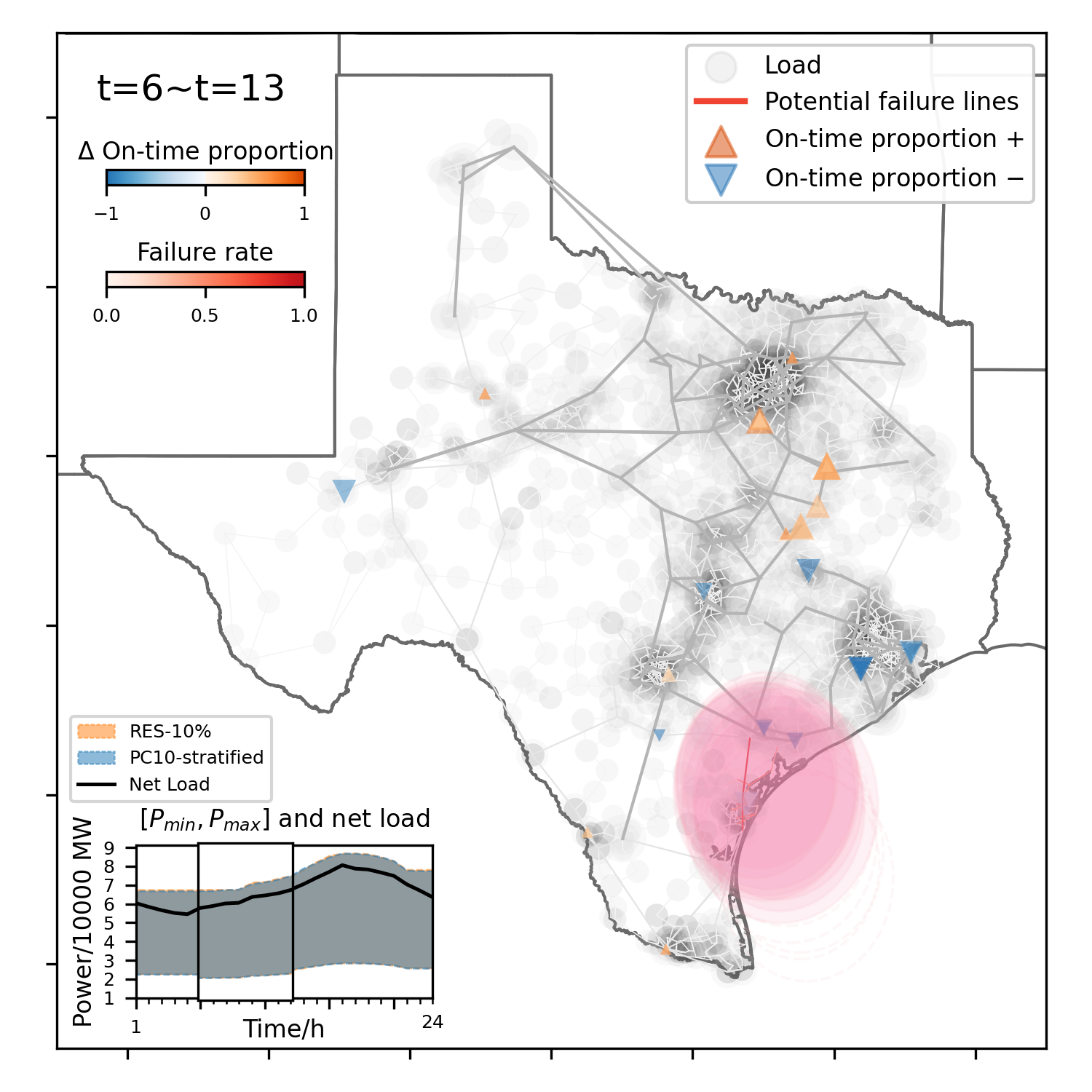}
    \caption{Thermal unit on-time proportion difference between preventive control and deterministic methods}
    \label{fig_on_off_diff_nDR}
\end{figure}

Having analyzed the cost characteristics of unit commitment decisions across different methods, we next explore how incorporating anticipated scenarios into preventive control affects unit commitment outcomes. Figure~\ref{fig_on_off_diff_nDR} compares the unit on-time proportion (i.e., the percentage of time a unit operates) between \emph{PC10-stratified} and \emph{RES-10 \%} \raa{(``+" means \emph{PC10-stratified}'s result has greater on-time proportion and ``-" vice versa)}. \raa{The size of triangle reflects the capacity of the unit.} Relative to the deterministic method, the preventive control method shuts down more units in hurricane-affected areas and activates more units farther north. This shift is driven by line congestion caused by transmission failures near the coast. Units in the hurricane-impacted region face congestion and must be shut down, with their loads being supplied by distant units unaffected by the hurricane.

\begin{figure}
    \centering
    \includegraphics[width=0.8\linewidth]{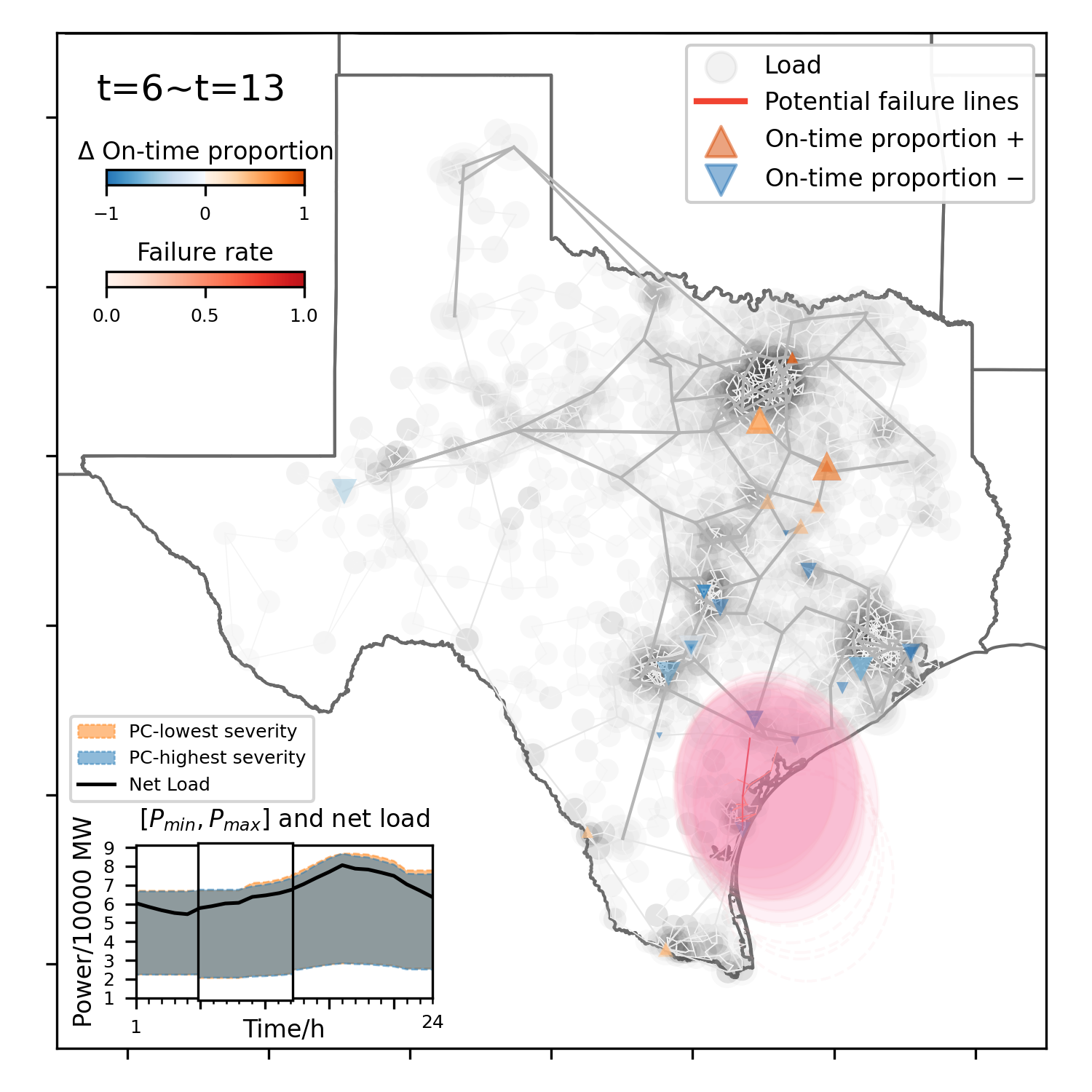}
    \caption{Thermal unit on-time proportion difference between preventive control methods considering the highest and lowest severity scenarios}
    \label{fig_on_off_diff_nDR_best_worst}
\end{figure}

To investigate these patterns further, we analyze unit commitment results from preventive control considering the most and least severe scenarios, as shown in Figure~\ref{fig_on_off_diff_nDR_best_worst}. The analysis shows that preventive control considering the least severe scenarios tends to activate more units in areas heavily impacted by extreme weather, which can exacerbate line congestion. Conversely, preventive control based on the most severe scenarios proactively shuts down these local units to prepare for extreme conditions. However, this precautionary approach may result in load curtailment when actual conditions are less severe.
In summary, preventive control strategies that account for more severe scenarios tend to shut down more units within and near hurricane-affected areas, anticipating line failures and potential congestion.

\begin{figure}
    \centering
    \includegraphics[width=0.8\linewidth]{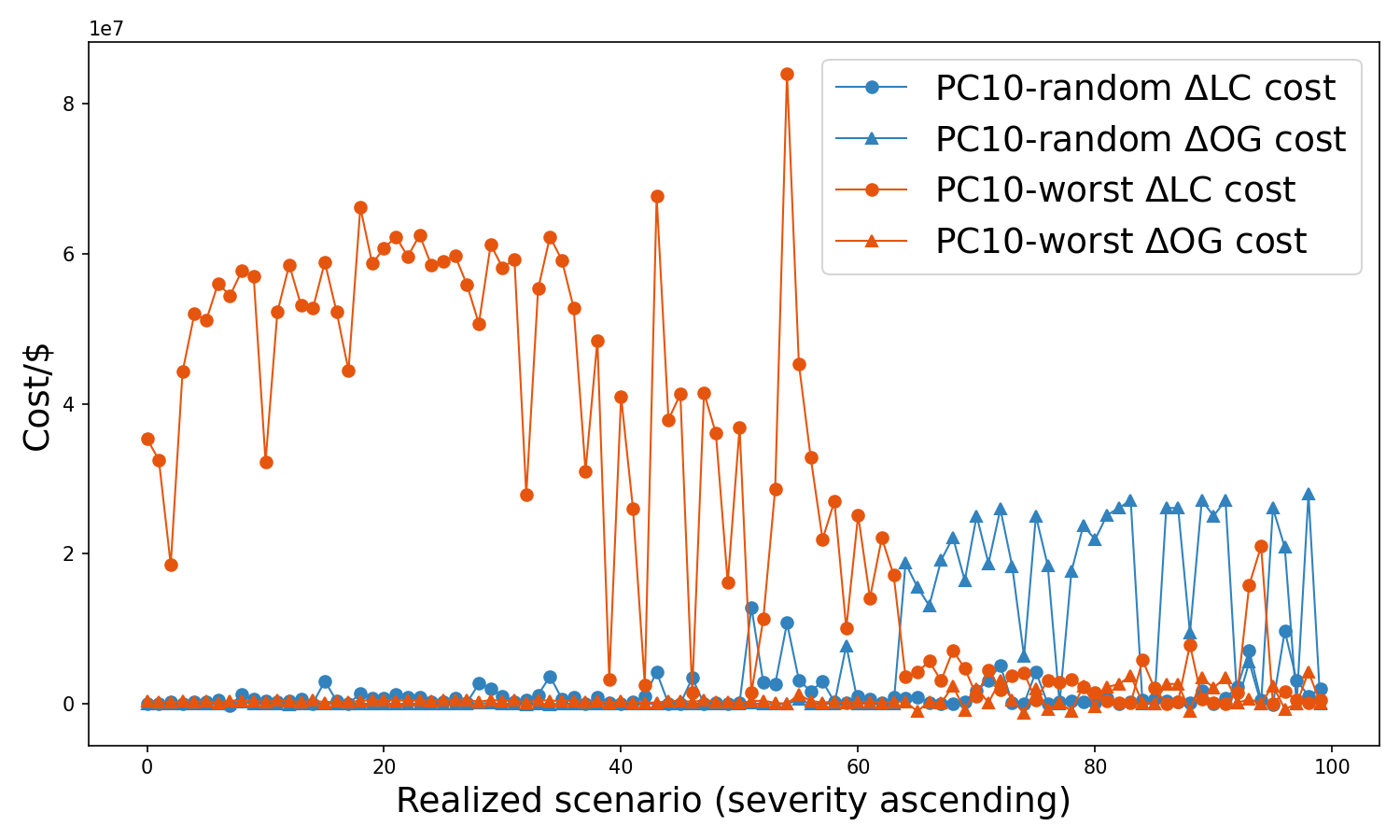}
    \caption{Load curtailment and over-generation costs of different methods in each realized scenario}
    \label{fig_cost_by_sce_cost_LC_SL_nDR}
\end{figure}

With respect to scenario impacts, preventive control explicitly incorporates risk by anticipating potential contingencies. By precisely adjusting unit start-up and shut-down schedules, it achieves a lower overall cost. On this cost baseline, the severity of scenarios \raa{$q(\mathbf{S}^\pi)$} considered in preventive control reflects different risk preferences, which in turn further strengthen or weaken changes in unit commitment patterns. This approach introduces a trade-off: higher load curtailment costs in milder realized scenarios, or higher over-generation costs in more severe scenarios. As the severity of considered scenarios \raa{$q(\mathbf{S}^\pi)$} increases, the primary source of additional costs shifts from over-generation to load curtailment, as shown in Figure~\ref{fig_cost_by_sce_cost_LC_SL_nDR}. Here, $\Delta$LC cost and $\Delta$OG cost represent the additional costs compared to the best outcome. \emph{PC10-worst}, which considers more severe scenarios, tends to incur higher load curtailment costs in milder realized scenarios, whereas \emph{PC10-random} which considers milder scenarios, tends to incur higher over-generation costs in more severe realized scenarios.

\subsection{Importance of Spatially Dependent Sampling}
\label{sec:importance_of_SDS}

The findings above demonstrate how the severity of selected scenarios \raa{$q(\mathbf{S}^\pi)$} affects the pattern and outcome of preventive control. Based on this general pattern, we now discuss how the spatially dependent sampling technique influences preventive control results \raa{compared with normal sampling method}. 

We generate 10,000 scenarios $\mathbf{S^{NP}}$ using normal sequential Monte Carlo simulations and select \emph{10-random} and \emph{10-worst} subsets $\mathbf{S^N}$ similarly. Preventive controls are then applied using these scenario subsets $\mathbf{S^N}$, and the results are tested against the previously mentioned scenario set $\mathbf{S^T}$. 
\raa{In Section~\ref{sec:importance_of_SDS}, we posit correlated ground truth to probe the bias introduced by independent sampling. This assumption follows the physical arguments in Sections~\ref{sec:two_component_failure} and \ref{sec:relevance_sampling_technique} and is used to reveal the implications of neglecting correlation.}

\begin{figure}
    \centering
    \includegraphics[width=0.8\linewidth]{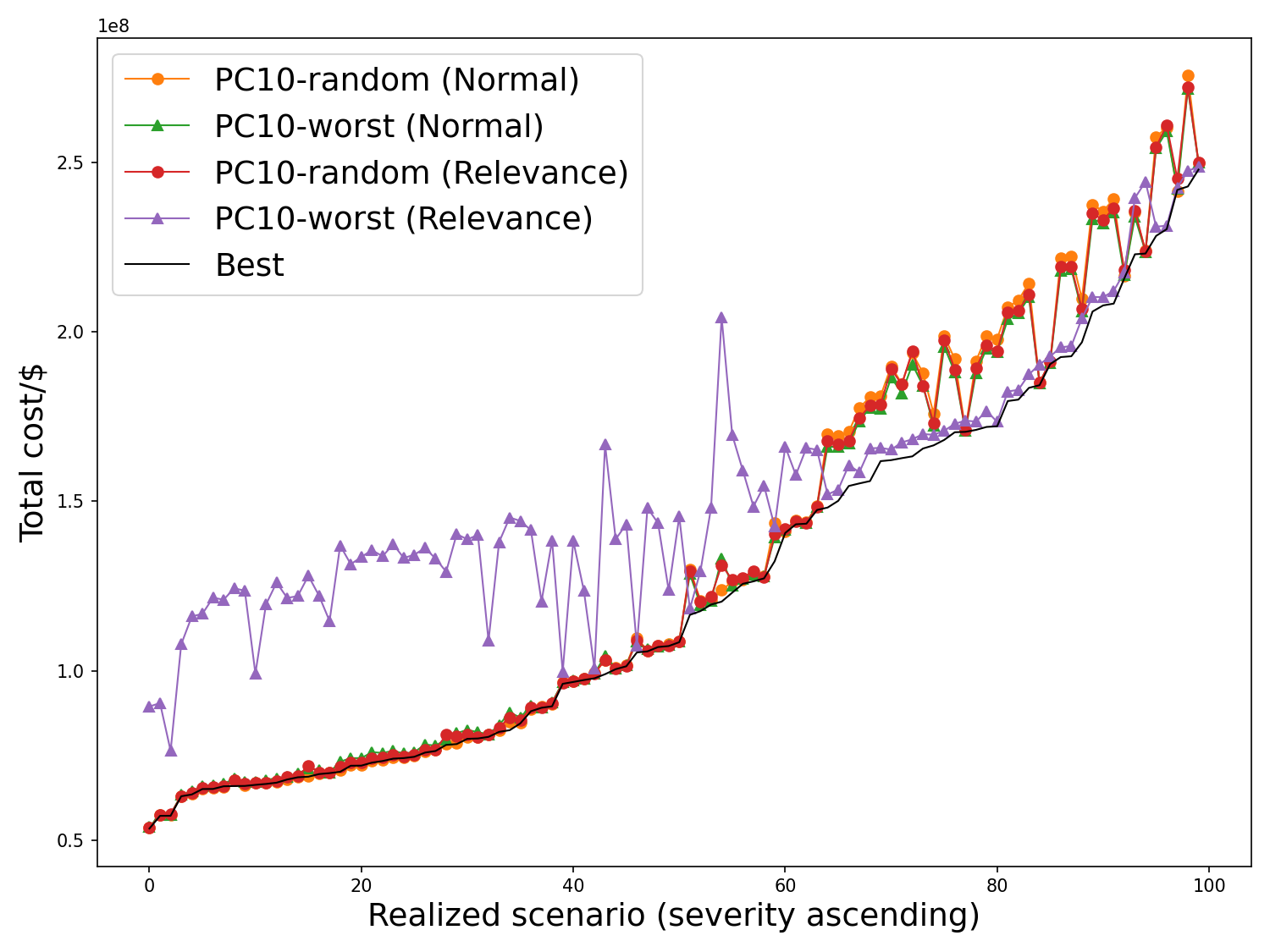}
    \caption{Total cost of different methods (considering scenarios generated by relevance and normal sampling) in each realized scenario}
    \label{fig_cost_by_sce_obj_DR_rel_nor}
\end{figure}

As shown in Figure~\ref{fig_cost_by_sce_obj_DR_rel_nor}, even when considering the \emph{10-worst} scenarios from the normal sampling set, preventive controls exhibit a stochastic optimization pattern that deviates from the ideal in severe cases.

Among the tested methods, only the preventive control method that considers \emph{10-worst} scenarios from the relevance sampling set $\mathbf{S^{RP}}$ achieves a robust optimization pattern, effectively managing severe cases. This is critical because normal sampling may overlook severe cases, particularly when component failures are correlated under extreme weather conditions. Ignoring these worst-case scenarios can render preventive controls ineffective during extreme events.

\begin{figure}
	\centering
	\includegraphics[width=0.75\linewidth]{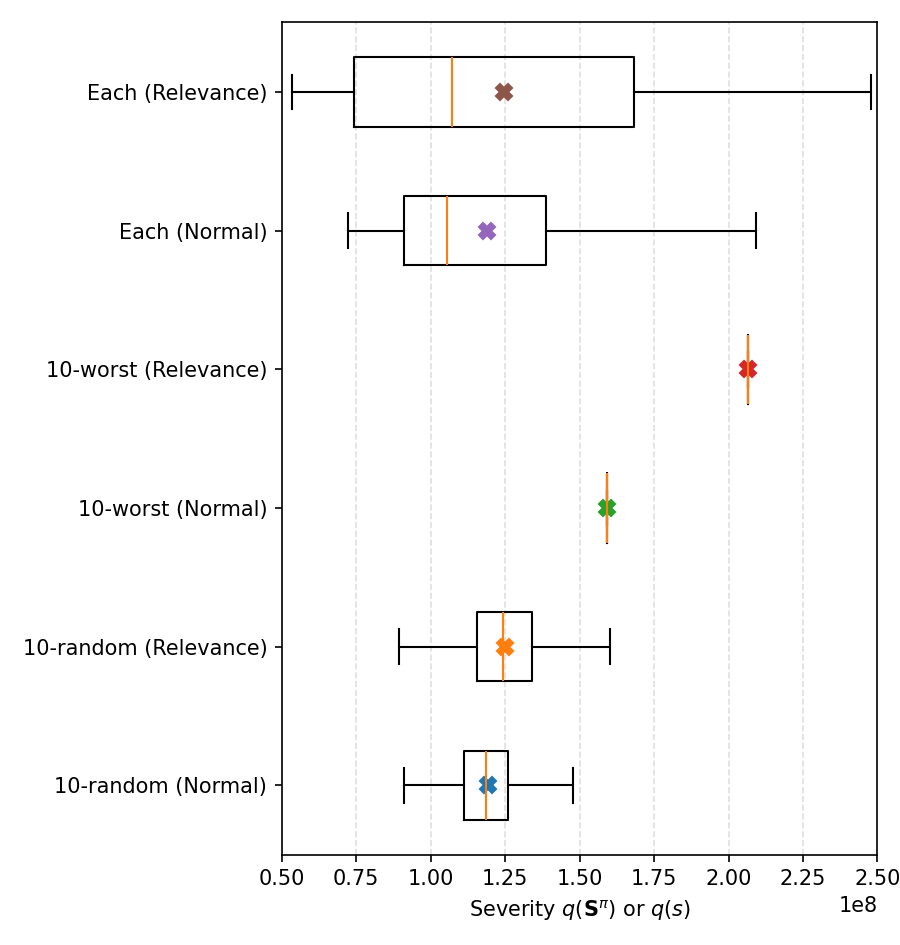}
	\caption{\raa{Severity distribution of $\mathbf{S}^\pi$ selected by each method from relevance and normal sampled scenarios}}
	\label{fig:severity_dist_ST_cross}
\end{figure}

\raa{
Figure~\ref{fig:severity_dist_ST_cross} reports $q(\mathbf{S}^\pi)$ for the selection rules compared in Figure~\ref{fig_cost_by_sce_obj_DR_rel_nor}. Because the scenarios in the normal sampling pool $\mathbf{S}^N$ are generally less severe, subsets chosen from $\mathbf{S}^N$ by the same rules (\emph{10-random} and \emph{10-worst}) also exhibit lower severity. The \emph{10-worst} subset from normal sampling still yields $q(\mathbf{S}^\pi)$ greater than the unbiased selections (random/stratified), but it is less risk-averse than the \emph{10-worst} subset from relevance sampling and therefore behaves more like an ``unbiased'' stochastic selection, rather than robust selection.
}

\section{Conclusion}

Preventive control is an effective strategy for power system operation during natural hazards. Its effectiveness hinges on the quality of the generated scenarios. Most previous works incorporating preventive control and scenario generation rely on sequential Monte Carlo simulations, which sample multiple components' failures independently at different time intervals. This paper considers the correlation of multiple component failures due to the correlated weather intensity random variables within the scope of a hurricane.

This paper discusses the correlation between component failures under varying levels of weather intensity correlation and proposes a spatially dependent sampling technique to sample correlated component failures. Using synthetic Texas grid data and the simulated Hurricane \textit{Harvey}, the patterns of generated scenarios are compared. Based on a stochastic unit commitment preventive control model, the impact of these scenarios on the preventive control outcomes is analyzed.

The major findings and conclusions are as follows:

\begin{enumerate}
    \item \raa{When weather intensity is strongly correlated with large enough uncertainty, regardless of the level of weather intensity, component failures are highly likely to be correlated.}
    \item The proposed spatially dependent sampling method can sample scenarios with longer tails in the distribution, effectively capturing more extreme scenarios.
    \item The preventive control method involves a trade-off between load curtailment costs in less severe scenarios and over-generation costs in more severe ones, depending on the severity of the considered scenarios.
    \item Ignoring the relevance of component failures can result in the omission of high-severity scenarios, thereby preventing robust optimization.
\end{enumerate}

\raa{Although we validated SDS using hurricanes and transmission line failures, the framework is hazard- and asset-agnostic. It applies to any extreme weather process that admits an intensity-field model and to any component class with a fragility curve based failure model. We introduced several simplifying assumptions to work within the constraints of limited weather uncertainty data. In practice, richer uncertainty characterizations, such as more detailed parameter errors in the meteorological model or a directly specified joint distribution of weather intensity, can be embedded in SDS to improve realism. The core of SDS is to model the spatial correlation of weather intensity random variables and to jointly sample component failures accordingly.}


\appendix
\section{Holland Wind Field Model}
\label{app:holland}
\raa{
This appendix summarizes the Holland wind field model used to compute surface wind speed at a target location $(\phi_2,\lambda_2)$ given a tropical cyclone centered at $(\phi_1,\lambda_1)$. Angles are in radians unless otherwise noted.
}

\raa{Let $\Delta\phi=\phi_2-\phi_1$ and $\Delta\lambda=\lambda_2-\lambda_1$. The great-circle (haversine) distance $r$ from the cyclone center to the target point is
\begin{align}
a &= \sin^2\!\left(\frac{\Delta\phi}{2}\right)
    + \cos\phi_1 \cos\phi_2 \sin^2\!\left(\frac{\Delta\lambda}{2}\right), \label{eq:hav-a}\\[2mm]
r &= 2 R \cdot \arcsin\!\big(\sqrt{a}\,\big), \qquad R = 6371~\text{km}. \label{eq:hav-r}
\end{align}
}

\raa{The azimuth (bearing) $\theta$ of the target relative to the cyclone center is
\begin{align}
\theta = \arctan\!\left(
\frac{\sin\Delta\lambda \,\cos\phi_2}{\cos\phi_1 \sin\phi_2 - \sin\phi_1 \cos\phi_2 \cos\Delta\lambda}
\right).
\label{eq:azimuth}
\end{align}
}

\raa{Let $S$ denote the cyclone translation speed and $\alpha$ its heading (clockwise from geographic north). Decompose $S$ into eastward and northward components:
\begin{align}
S_{\text{east}} &= S \sin\alpha, \label{eq:Seast}\\
S_{\text{north}} &= S \cos\alpha. \label{eq:Snorth}
\end{align}
}

\raa{The Holland gradient wind speed at range $r$ is
\begin{equation}
V_g(r) = \sqrt{ \frac{B (P_n - P_c)}{\rho}\,
e^{-\left(\frac{R_{\text{max}}}{r}\right)^{\!B}}
\left(\frac{R_{\text{max}}}{r}\right)^{\!B}
+ \left(\frac{r f}{2}\right)^{\!2} }
\;-\; \frac{r f}{2},
\label{eq:Vg}
\end{equation}
where $P_n$ is the environmental pressure, $P_c$ is the central pressure, $\rho$ is air density, $R_{\text{max}}$ is the radius of maximum wind, $B$ is the Holland shape parameter, and $f$ is the Coriolis parameter
\begin{equation}
f = 2\Omega \sin\phi, \qquad \Omega \approx 7.292\times10^{-5}~\text{rad s}^{-1}.
\label{eq:coriolis}
\end{equation}
In \eqref{eq:coriolis}, $\phi$ is the local latitude (e.g., take $\phi=\phi_2$).
}

\raa{Assuming cyclonic flow that is tangential to circles around the center, the gradient-wind components (east, north) at the target are
\begin{align}
u_g &= V_g(r)\,\cos\!\big(\theta + 90^{\circ}\big), \label{eq:ug}\\
v_g &= V_g(r)\,\sin\!\big(\theta + 90^{\circ}\big). \label{eq:vg}
\end{align}
The total wind components add the translation:
\begin{align}
u_{\text{total}} &= u_g + S_{\text{east}}, \label{eq:utot}\\
v_{\text{total}} &= v_g + S_{\text{north}}. \label{eq:vtot}
\end{align}
The resulting surface wind speed is
\begin{equation}
V_{\text{total}} \;=\; f^{\text{Holland}}_i\!\left(P_c, R_{\text{max}}, B, \phi_1, \lambda_1, S, \alpha\right)
\;=\; \sqrt{u_{\text{total}}^{\,2} + v_{\text{total}}^{\,2}}.
\label{eq:Vtotal}
\end{equation}
}

\section{Linearization Assumption Validation}

\raa{
We define the \emph{linearity deviation} index
\[
d(\boldsymbol{\theta},\Delta\boldsymbol{\theta}) =
\left|
\frac{f^W(\boldsymbol{\theta}+\Delta\boldsymbol{\theta})-f^W(\boldsymbol{\theta})
-\Delta\boldsymbol{\theta}^{\top}\nabla_{\boldsymbol{\theta}} f^W(\boldsymbol{\theta})}
{\Delta\boldsymbol{\theta}^{\top}\nabla_{\boldsymbol{\theta}} f^W(\boldsymbol{\theta})}
\right|,
\]
which quantifies the relative mismatch between the actual change
$\Delta f^W(\boldsymbol{\theta})$ and its first-order linear approximation
$\Delta\boldsymbol{\theta}^{\top}\nabla_{\boldsymbol{\theta}} f^W(\boldsymbol{\theta})$.
}

\begin{figure}
  \centering
  \includegraphics[width=0.75\linewidth]{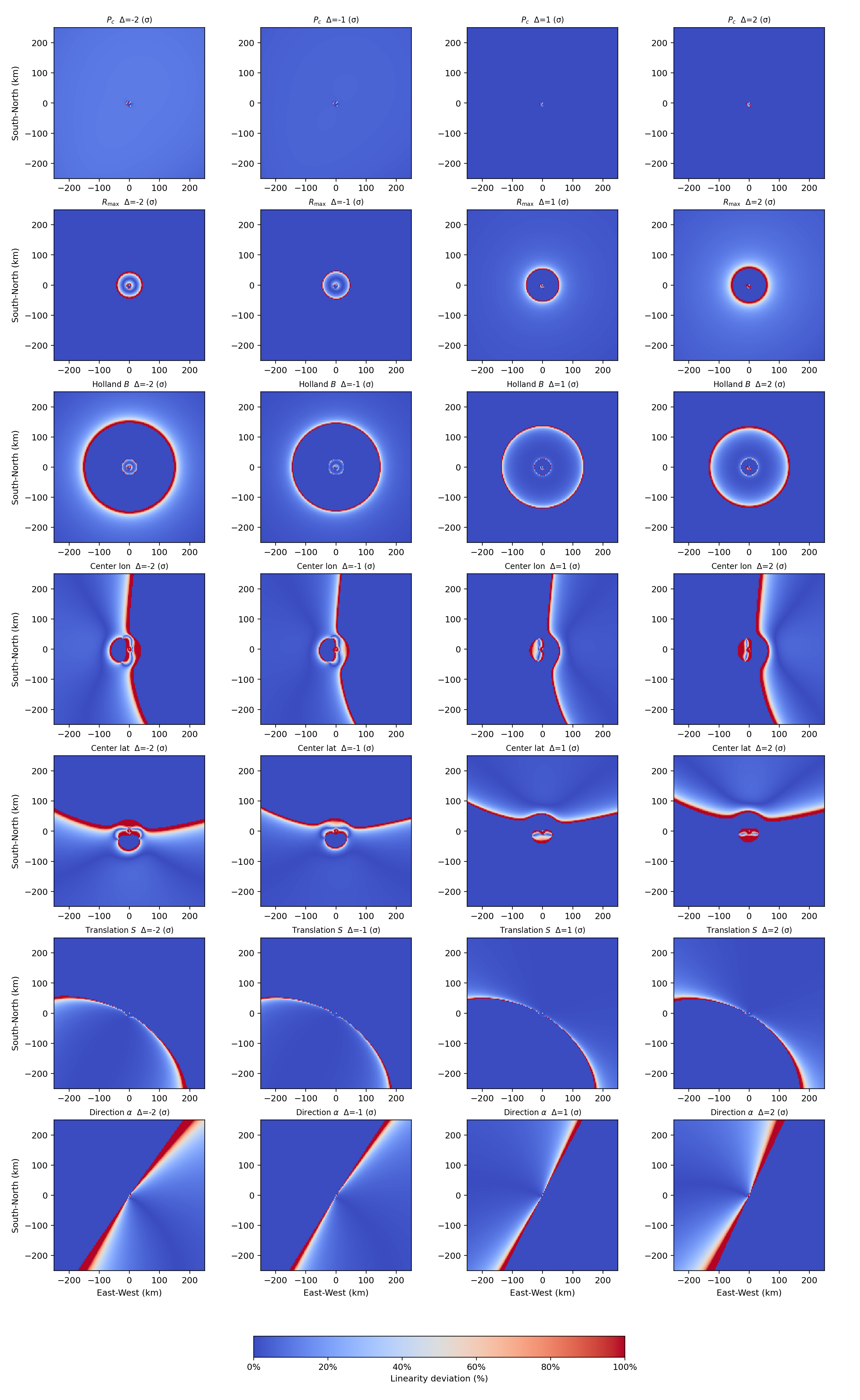}
  \caption{\raa{Linearity deviation for locations around the hurricane.}}
  \label{fig:linearity_deviation}
\end{figure}

\raa{
We consider a hurricane characterized by
$P_c=975\,\mathrm{hPa}$, $R_{\max}=50\,\mathrm{km}$, $B=1.3$,
$\phi_1=22.3^\circ$, $\lambda_1=-96^\circ$, $S=5\,\mathrm{m/s}$, and $\alpha=-30^\circ$.
This is an intense hurricane with maximum wind speed around 50~m/s.
Parameter uncertainties are set to
$\sigma(P_c)=10\,\mathrm{hPa}$, $\sigma(R_{\max})=5\,\mathrm{km}$,
$\sigma(B)=0.05$, $\sigma(\phi_1)=0.1^\circ$, $\sigma(\lambda_1)=0.1^\circ$,
$\sigma(S)=0.5\,\mathrm{m/s}$, and $\sigma(\alpha)=5^\circ$.
We construct a $500\,\mathrm{km}\times 500\,\mathrm{km}$ mesh centered on the storm and,
for each grid location, perturb each of the seven parameters by
$\{-2\sigma,-\sigma,+\sigma,+2\sigma\}$ (one at a time) to compute the
\emph{linearity deviation}. The resulting heatmaps are shown in
Figure~\ref{fig:linearity_deviation}.
}

\raa{Linearity deviation is small across most locations in the figure; only a few narrow, band-shaped regions exhibit large errors under the linearization. This supports the validity of using a linearized expansion.}




\end{document}